\documentclass[aps,prd,twocolumn,floats,floatfix,showpacs,amssymb,prd,superscriptaddress,nofootinbib]{revtex4-1}
\usepackage{graphicx, epsfig, amssymb} %include figure files
\usepackage{amsmath, amsfonts}
\usepackage{bm}

\usepackage[usenames]{xcolor}
\definecolor{linkcolor}{rgb}{0.0,0.3,0.5}
\usepackage[unicode,colorlinks=true,citecolor=linkcolor,linkcolor=linkcolor,urlcolor=linkcolor]{hyperref}
\usepackage[caption=false]{subfig}

\def\be{\begin{equation}}
\def\ee{\end{equation}}
\newcommand{\beq}{\begin{eqnarray}}
\newcommand{\eeq}{\end{eqnarray}} 
\newcommand{\ba}{\begin{align}}
\newcommand{\ea}{\end{align}}

\begin{document}

\title{\large Black-hole head-on collisions in higher dimensions}

\author{William G. Cook}\email{wc259@cam.ac.uk}
\affiliation{Department of Applied Mathematics and Theoretical Physics,
Centre for Mathematical Sciences, University of Cambridge,
Wilberforce Road, Cambridge CB3 0WA, United Kingdom}
\author{Ulrich Sperhake}\email{u.sperhake@damtp.cam.ac.uk}
\affiliation{Department of Applied Mathematics and Theoretical Physics,
Centre for Mathematical Sciences, University of Cambridge,
Wilberforce Road, Cambridge CB3 0WA, UK}
\affiliation{TAPIR 350-17, California Institute of Technology, 1200 E California
Boulevard, Pasadena, California 91125, USA}
\affiliation{Department of Physics and Astronomy, University of Mississippi,
University, Mississippi 38677, USA}
\author{Emanuele Berti}\email{eberti@olemiss.edu}
\affiliation{Department of Physics and Astronomy, University of Mississippi,
University, Mississippi 38677, USA}
\affiliation{CENTRA, Departamento de F\'{\i}sica, Instituto Superior
T\'ecnico, Universidade de Lisboa,
Avenida Rovisco Pais 1, 1049 Lisboa, Portugal}
\author{Vitor Cardoso}\email{vitor.cardoso@ist.utl.pt}
\affiliation{CENTRA, Departamento de F\'{\i}sica, Instituto Superior
T\'ecnico, Universidade de Lisboa,
Avenida Rovisco Pais 1, 1049 Lisboa, Portugal}
\affiliation{Perimeter Institute for Theoretical Physics, Waterloo, Ontario N2L 2Y5, Canada}

\begin{abstract}
  The collision of black holes and the emission of gravitational
  radiation in higher-dimensional spacetimes are of interest in
  various research areas, including the gauge-gravity duality, the TeV
  gravity scenarios evoked for the explanation of the hierarchy
  problem, and the large-dimensionality limit of general
  relativity. We present numerical simulations of head-on collisions
  of nonspinning, unequal-mass black holes starting from rest in
  general relativity with $4\leq D\leq 10$ spacetime dimensions.  We
  compare the energy and linear momentum radiated in gravitational
  waves with perturbative predictions in the extreme mass ratio limit,
  demonstrating the strength and limitations of black-hole
  perturbation theory in this context.
\end{abstract}

%04.25.D-     Numerical relativity
%04.25.dg     Numerical studies of black holes and black-hole binaries
%04.50.-h     Higher-dimensional gravity and other theories of gravity
%04.50.Gh     Higher-dimensional black holes, black strings, and related objects

\maketitle

%%%%%%%%%%%%%%%%%%%%%%%%%%%%
\section{Introduction}
\label{sec:intro}
%%%%%%%%%%%%%%%%%%%%%%%%%%%
The study of higher-dimensional spacetimes dates back at least one
hundred years to the seminal attempts by Kaluza and Klein to unify
gravitation and electromagnetism.  Higher dimensional arenas would
resurface several times over the next decades, either in the context
of specific physical theories, such as string theory, or theories
which can be embedded into it. One particularly intriguing example is
the class of TeV-scale gravity theories, which propose to lower the
fundamental Planck scale by diluting gravity in a large number of
dimensions~\cite{Antoniadis:1990ew,Antoniadis:1990ew,ArkaniHamed:1998rs,Antoniadis:1998ig,
  Randall:1999ee,Randall:1999vf}. These proposals suggest that
dynamical processes involving higher-dimensional black holes (BHs) may
be relevant for understanding the physics under experimental scrutiny
at particle colliders, such as the Large Hadron Collider (LHC).  In
these scenarios, BH production would become possible at much lower
energies than the four-dimensional Planck scale $10^{19}\,{\rm GeV}$,
a possibility that remains interesting despite the robust constraints
at current LHC energies~\cite{Aaboud:2017yvp,Sirunyan:2017anm}.
In this framework, the understanding of BH dynamics and gravitational
radiation emitted during high-energy encounters is
fundamental~\cite{Cardoso:2014uka}.

Higher-dimensional spacetimes have also been used as a purely
mathematical construct, where the number $D$ of spacetime dimensions
is regarded as just one other parameter to be varied. Emparan and
collaborators~\cite{Emparan:2013moa,Emparan:2015hwa,Emparan:2015gva,Emparan:2014cia}
have recently added an elegant twist to this aspect of
higher-dimensional spacetimes by focusing on the large-$D$ limit. They
showed that the physics of four-dimensional spacetimes can be
recovered to good precision from a large-$D$ expansion, and that the
large-$D$ limit offers precious physical insight into the nature of
classical and quantum gravity in arbitrary dimensions.

The purpose of this work is to extend previous results on the
low-energy collision of BHs to higher dimensions.  This effort was
started a few years ago~\cite{Witek:2010az,Witek:2014mha}, but a
combination of gauge issues and difficulties in the regularization of
variables in the dimensional reduction generated numerical
instabilities, restricting all binary BH simulations to $D\le 6$
spacetime dimensions.  Building on earlier work
\cite{Pretorius:2004jg,Yoshino:2009xp} on the so-called {\em modified
  cartoon method}, Refs.~\cite{Cook:2016soy,Cook:2016qnt} reported
considerable progress in overcoming stability limitations and in the
numerical extraction of gravitational waves (GWs) in
higher-dimensional spacetimes.  Using the methods developed therein,
we present new results for the collision of unequal-mass BH binaries
in $D=4,5,\dots,10$ dimensions, and compare these with perturbative
predictions. We expect our results to also allow for making contact
with the large-$D$ regime studied by Emparan and collaborators.

%%%%%%%%%%%%%%%%%%%%%%%%%%%%%%%%%%%%%
\section{Modeling framework}
\label{sec:modelling}
%%%%%%%%%%%%%%%%%%%%%%%%%%%%%%%%%%%%%
The physical scenario we consider in this work consists of two
$D$-dimensional, nonspinning BHs with masses $M_1$ and $M_2\le M_1$
initially at rest, which then collide head-on under their
gravitational attraction and merge into a single BH.  The
gravitational radiation released during the encounter of the two BHs,
and its total energy and momentum
in particular, is the key diagnostic quantity we
wish to extract from our calculations.  For this purpose, we employ
two techniques: (i) a perturbative point-particle (PP) approximation,
and (ii) numerical relativity simulations assuming $SO(D-3)$
isometry. In this section we review these two methods in turn.

%%%%%%%%%%%%%%%%%%%%%%%%%%%%%%%%%%%%%
\subsection{Point-particle calculations}
\label{sec:PP}
%%%%%%%%%%%%%%%%%%%%%%%%%%%%%%%%%%%%%

The first attempt at understanding this process considers a somewhat
restricted parameter space: one of the BHs is much more massive than
the other, i.e. $q\equiv M_2/M_1\ll 1$ or
\begin{equation}
\eta \equiv \frac{M_1M_2}{(M_1+M_2)^2}
        = \frac{q}{(1+q)^2} \ll 1\,,
\end{equation}
where $\eta$ is the symmetric mass ratio.  The smaller, lighter BH is
then approximated as a structureless PP, moving on a geodesic of the
background spacetime described by the massive BH, while generating a
stress-energy tensor which perturbs it. This scheme is also sometimes
known as the PP approximation. In such a framework, the resulting
equations to solve are just linearized versions of the Einstein
equations, expanded around a BH-background
spacetime~\cite{Regge:1957td,Zerilli:1971wd,Davis:1971gg,Kodama:2003jz,Berti:2003si,Berti:2010gx}. When
the massive BH is nonspinning, the equations reduce to a single
ordinary differential equation sourced by the smaller BH (the PP). In
this scheme, to leading order, the total energy
$E_{\rm rad}\propto q^2$. The exact coefficient was computed in
Refs.~\cite{Davis:1971gg,Berti:2003si,Berti:2010gx} for particles
falling radially into the BH.

\begin{table}[t!]
\centering
\begin{tabular}{c|ccccc}
\hline
\hline
$D$                         &  4          & 5      & 6      & 7      & 8\\
\hline
$\frac{E_{\rm rad}}{q^2M}$&  0.0104     & 0.0165 & 0.0202 & 0.0231 & $0.0292$\\
\hline
\hline
\end{tabular}    
\caption{\footnotesize Energy radiated in GWs when a small BH of mass $qM_1,\,q\ll1$ falls from rest at infinity into a $D$-dimensional BH of mass $M_1$.
}
\label{tab:summaryPP}
\end{table}

Table~\ref{tab:summaryPP} summarizes those results for different
spacetime dimensions. Note that the proportionality coefficient
increases with spacetime dimension at large $D$. An extrapolation of
these results suggests that the perturbative PP calculation should
cease to be valid at sufficiently large $D$, since the radiation
ultimately becomes too large and the geodesic approximation breaks
down: cf. the discussion around Fig.~1 of \cite{Berti:2010gx}. Thus,
even within the PP approximation, we identify the need to solve the
full, nonlinear Einstein equations at large $D$.

%%%%%%%%%%%%%%%%%%%%%%%%%%%%%
\subsection{Numerical framework}
\label{sec:NR}
%%%%%%%%%%%%%%%%%%%%%%%%%%%%%
The only presently known method to solve the Einstein equations in the
dynamic and fully nonlinear regime is to use numerical tools on
supercomputers: see
e.g.~\cite{Pretorius:2007nq,Centrella:2010mx,Sperhake:2014wpa}.  In
higher dimensions, however, the computational cost increases rapidly
with $D$. To achieve sufficient resolution of all relevant scales, typical grid sizes in our simulations have $\mathcal{O}({10^2})$ grid points in each dimension, hence the computational cost increases approximately by this factor for each
increment in $D$, making it impractical to consider arbitrary $D-1$
dimensional spatial grids.  Many physical scenarios of current
interest, however, involve degrees of symmetry in the extra dimensions
that facilitate a reduction of the effective computational domain to
three or fewer spatial dimensions, as handled in traditional numerical
relativity. The physical effects of the extra dimensions are then
encapsulated in a set of additional fields on the effective domain.
Several
approaches to achieve such a dimensional reduction have been
implemented in the literature: see
e.g.~\cite{Pretorius:2004jg,Zilhao:2010sr,Sorkin:2009wh,Yoshino:2011zz}.
Here, we use the modified cartoon method in the form
detailed in \cite{Cook:2016soy}, which describes a $D$-dimensional
spacetime with $SO(D-3)$ isometry.

Specifically, we use the {\sc lean}
code~\cite{Sperhake:2006cy,Sperhake:2007gu}, originally developed for
BH simulations in $D=4$ dimensions and upgraded to general $D$
spacetime dimension with $SO(D-3)$ isometry
in~\cite{Zilhao:2010sr,Cook:2016soy}.  We start our simulations with
the $D$ dimensional generalization of Brill-Lindquist
\cite{Brill:1963yv} data in Cartesian coordinates $X^I$ (Capital Latin
indices $I,\,J,\,\ldots$ cover the range $1,\,\ldots,\,D-1$, while
lower case Latin indices $i, j, \ldots$ cover the range $1, 2, 3$),
\begin{eqnarray}
  &&\gamma_{IJ} = \psi^{4/(D-3)} \delta_{IJ}\,,~~~~~~~~~~K_{IJ}=0\,,
        \nonumber \\[10pt]
  &&\psi = 1+\sum_{\mathcal{N}} \frac{\mu_{\mathcal{N}}}
        {4\left[ \sum_K (X^K-X^K_{\mathcal{N}})^2\right]^{(D-3)/2}}\,,
  \label{eq:BLD}
\end{eqnarray}
where $\gamma_{IJ}$ and $K_{IJ}$ are the spatial metric and extrinsic
curvature of the Arnowitt-Deser-Misner (ADM)
\cite{Arnowitt:1962hi} formalism and we set $G=c=1$.
The index $\mathcal{N}$ labels the individual BHs and, in our case,
always extends over the range $\mathcal{N}=1,\,2$.
These data are evolved in time with
the Baumgarte-Shapiro-Shibata-Nakamura
\cite{Shibata:1995we,Baumgarte:1998te} formulation of the Einstein
equations, combined with the moving puncture
\cite{Baker:2005vv,Campanelli:2005dd} gauge and Berger-Oliger
mesh refinement provided by {\sc carpet}
\cite{Schnetter:2003rb,Carpetweb} as part of the {\sc cactus}
computational toolkit \cite{Allen:1999,Cactusweb}.
In order to calculate the GW signal, we compute the
higher-dimensional Weyl scalars, as detailed in
\cite{Godazgar:2012zq,Cook:2016qnt,Cook:inprep}.
For comparison and to determine the
contributions of the individual multipoles, we also extract waveforms
calculated with the perturbative Kodama-Ishibashi approach
\cite{Kodama:2003jz,Ishibashi:2011ws} as detailed in
\cite{Witek:2010xi}.

Compared with previous simulations of BH collisions in higher dimensions,
we have implemented two changes we find necessary to achieve accurate
and stable evolutions. First, we evolve the lapse function $\alpha$
according to
\begin{equation}
  \partial_t \alpha = \beta^i \partial_i \alpha - c_1 \alpha K^{c_2}\,,
\end{equation}
where $\beta^i$ is the shift vector and $K$ the trace of the extrinsic
curvature; the slicing condition typically used in moving puncture
simulations is recovered for $c_1=2$, $c_2=1$ -- cf.~Eq.~(11) in
\cite{Sperhake:2006cy} -- but here we vary these parameters in the
ranges $2\le c_1\le 10$ and $1\le c_2\le 1.5$. The exact values vary
from configuration to configuration and have been determined
empirically. The second modification is an approximately linear
reduction of the Courant factor $\Delta t/\Delta x$ as a function of
$D$ from $0.5$ in $D=4$ to $0.03$ in $D=10$. We shall see in
Fig.~\ref{fig:delta} and its discussion in Sec.~\ref{sec:results_EM}
that the merger becomes an increasingly instantaneous event with an
ever sharper burst in radiation as we increase $D$. We believe the
necessity of reducing the Courant factor to arise from this increasing
demand for time resolution around merger.

%%%%%%%%%%%%%%%%%%%
\section{Results}
\label{sec:results}
%%%%%%%%%%%%%%%%%%%
In Schwarzschild coordinates, a nonrotating
$D$-dimensional BH with ADM
mass $M$ has a horizon or Schwarzschild radius given by
\begin{equation}
  R_{S}^{D-3} = \frac{16\pi M}{(D-2)\mathcal{A}_{D-2}}\,,
  \label{eq:RSMADM}
\end{equation}
where
$\mathcal{A}_{D-2} = 2\pi^{(D-1)/2}/\Gamma\left(\frac{D-1}{2}\right)$.
Note that $R_S$ is related to the mass parameter $\mu$ of
the single BH version of Eq.~(\ref{eq:BLD}) by $\mu=R_S^{D-3}$.
In consequence of Eq.~(\ref{eq:RSMADM}),
mass and length do not have the same physical
dimensions unless $D=4$. Henceforth, we measure energy in units of the
ADM mass $M$ of the spacetime under consideration, and we measure
length and time in units of the Schwarzschild radius $R_S$ associated
with this ADM mass according to Eq.~(\ref{eq:RSMADM}).
%%%%%%%%%%%%%%%%%%%%%%%%%%%%%%%%%%%%%%%%%%%%%%%%%%%
\subsection{Numerical uncertainties}
%%%%%%%%%%%%%%%%%%%%%%%%%%%%%%%%%%%%%%%%%%%%%%%%%%%
Our numerical relativity
results for the GW energy released in head-on collisions
of BHs are affected by the following uncertainties:  \\[10pt]
\textit{\textbf{Discretization error.}} We
estimate the error due to finite grid resolution by
studying a head-on
collision of two BHs in $D=8$ dimensions with mass ratio
$q=1/20$. We use a computational grid composed of 8 nested refinement
levels, 2 inner boxes initially centered on the individual holes,
and 6 outer levels centered on the origin. The grid spacing around the
BHs is $h_1=R_S/113$, $h_2=R_S/129$ and $h_3=R_S/145$, respectively,
\begin{figure}[tb]
  \includegraphics[width=0.5\textwidth,clip]{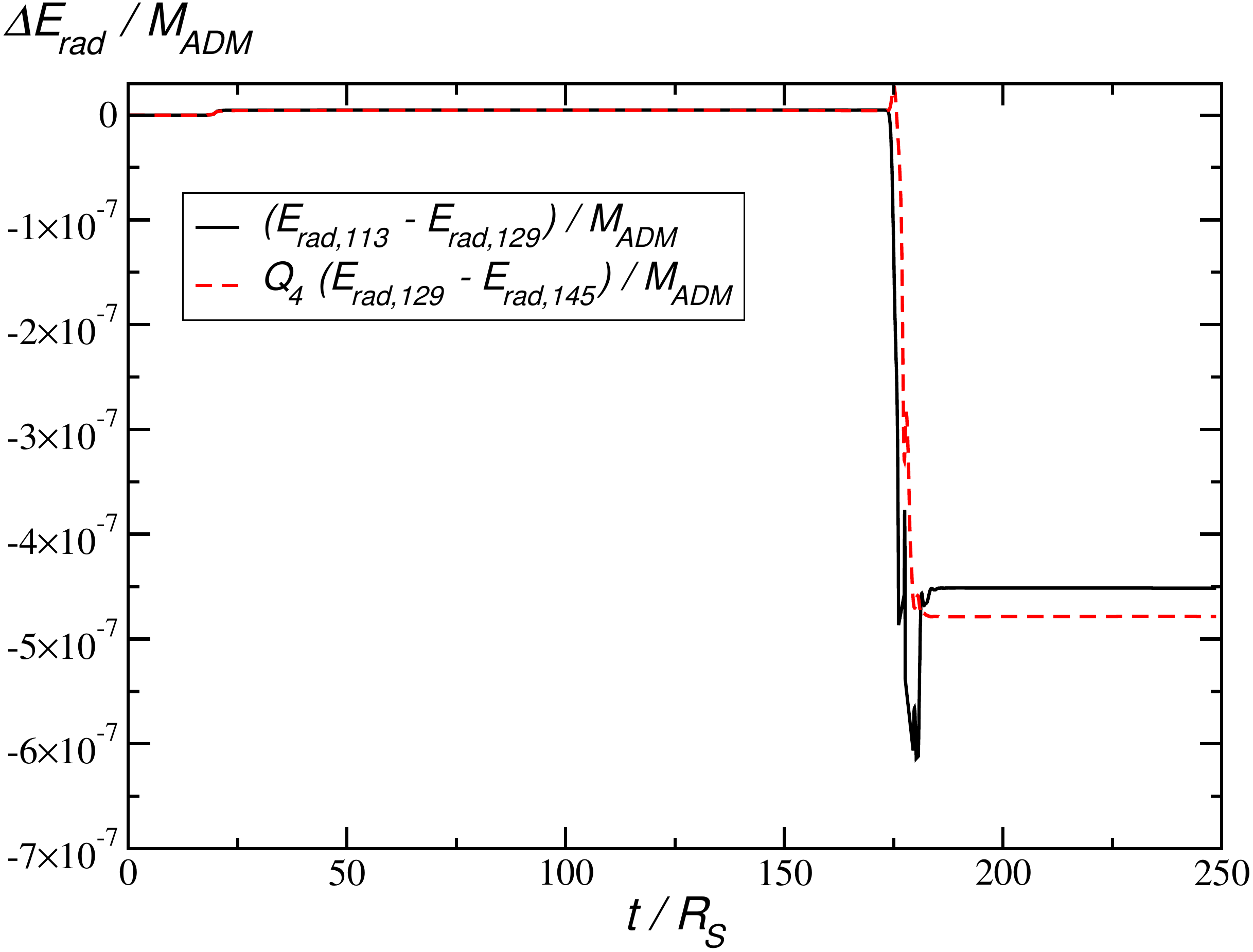}
  \caption{Convergence plot for the radiated energy $E_{\rm rad}$
           extracted from a $q=1/20$ head-on collision in $D=8$
           at $40~R_S$ as a function of time for
           grid spacing $h_1=R_S/113$, $h_2=R_S/129$ and
           $h_3=R_S/145$. The difference between the high and
           medium resolution simulations has been scaled by a
           factor $Q_4=1.88$ expected for fourth-order convergence
           and agrees well with the difference of the coarse and
           medium resolution energies.}
  \label{fig:conv}
\end{figure}
in our three simulations for checking convergence, and increases by a
factor 2 on each consecutive outer level.  The radiated energy as a
function of time is extracted at $40~R_S$, where the grid resolution
is $H_i=32\,h_i$ for the three runs $i=1,\,2,\,3$. The difference
between the high and medium resolution runs is compared with that
between the medium and coarse resolution runs in Fig.~\ref{fig:conv}.
Multiplying the former by a factor $Q_4=1.88$ (as expected for the
fourth-order discretization of the code) yields good agreement between the two
curves, and using the according Richardson-extrapolated result gives
an error estimate of $3\,\%$ for the medium resolution simulation,
which is closest to our set of production runs in terms of resolution
around the smaller BH and in the wave extraction zone.

We have analyzed several other configurations (including the collision
in $D=10$ dimensions) and find the discretization error to
mildly increase with mass ratio and dimensionality $D$,
from about $1\,\%$ for $q= 1$, $D=5,\,6$ to about
$4\,\%$ for $q=1$, $D=10$ and about $5\,\%$ for
$q\ll 1$, $D=8$.
\\[10pt]
\textit{\textbf{Finite extraction radius.}}
The computational domain used in our simulations is of finite extent,
about $200~R_S$ for the runs analyzed here, so that we cannot
extract the GW signal at infinity. Instead we
use finite radii and estimate the uncertainty incurred through
this process by fitting the total radiated energy using a
polynomial in $1/r_{\rm ex}$,
\begin{equation}
  E_{\rm rad}(r_{\rm ex}) = E_{\rm rad} + \frac{a}{r_{\rm ex}}
        +\mathcal{O}\left(\frac{1}{r_{\rm ex}^2}\right)\,,
\end{equation}
where $a$ is a parameter determined through fitting and $E_{\rm rad}$
is the estimate for the radiated energy extracted at infinity.
We then take the extrapolated value at infinity as our result,
and its difference from the largest numerical extraction radius as
the uncertainty estimate.
Applying this procedure yields a fractional error ranging from
about $0.4~\%$ for all equal-mass collisions
to about $4\,\%$ for configurations with $q\ll 1$.
\\[10pt]
\textit{\textbf{Spurious waves.}} Initial data of the type used here
typically contain a small amount of unphysical GWs colloquially
referred to as ``junk radiation''. The amount of unphysical radiation
depends on the initial separation of the BHs (vanishing in the limit
of infinite distance) and on the number of dimensions. As in
Ref.~\cite{Cook:2016qnt}, we find the amount of spurious radiation to
be orders of magnitude below the errors due to discretization and
extraction radius. We attribute this to the rapid falloff of gravity
in higher dimensions, so that the constituent BHs of the
Brill-Lindquist data are almost in isolation
even for relatively small coordinate separations.
We have noticed, however, that spurious radiation is more
prominent in the Kodama-Ishibashi modes as compared with the
results based on the Weyl scalars. We cannot account for the
precise causes for the seemingly superior behavior of the Weyl scalars,
but we note that similar findings have been reported for
the $D=4$ case in \cite{Reisswig:2010cd}.  
\\[10pt]
\textit{\textbf{Initial separation.}} The head-on collisions performed
here start from finite initial separation of the BHs, while the
idealized scenario considers two BHs falling in from infinity.  By
varying the initial separation for several collisions in $D=5$ and
$D=6$ we estimated the difference in $E_{\rm rad}$ due to the initial
separation and, as for the junk radiation above, we found that the
differences are well below the numerical error budget. Again, we
attribute this observation to the rapid falloff of the gravitational
attraction for large $D$,
leading to a prolonged but nearly stationary infall phase
followed by an almost instantaneous merger that
generates nearly all of the radiation. \\[10pt]
In summary, our error is dominated by discretization and use of finite
extraction radii. It ranges from about $1.5\,\%$ for comparable mass
collisions in low $D$ to about $9\,\%$ for $q\ll 1$ in $D=8$.  For the
gravitational recoil, we find similar significance of the individual
error contributions, but overall larger uncertainties by about a
factor 4. We attribute these larger uncertainties to the fact that the
recoil arises from asymmetries in GW emission, and in this sense it is
a weaker, differential effect.

\begin{figure}[tb]
  \includegraphics[width=0.5\textwidth,clip]{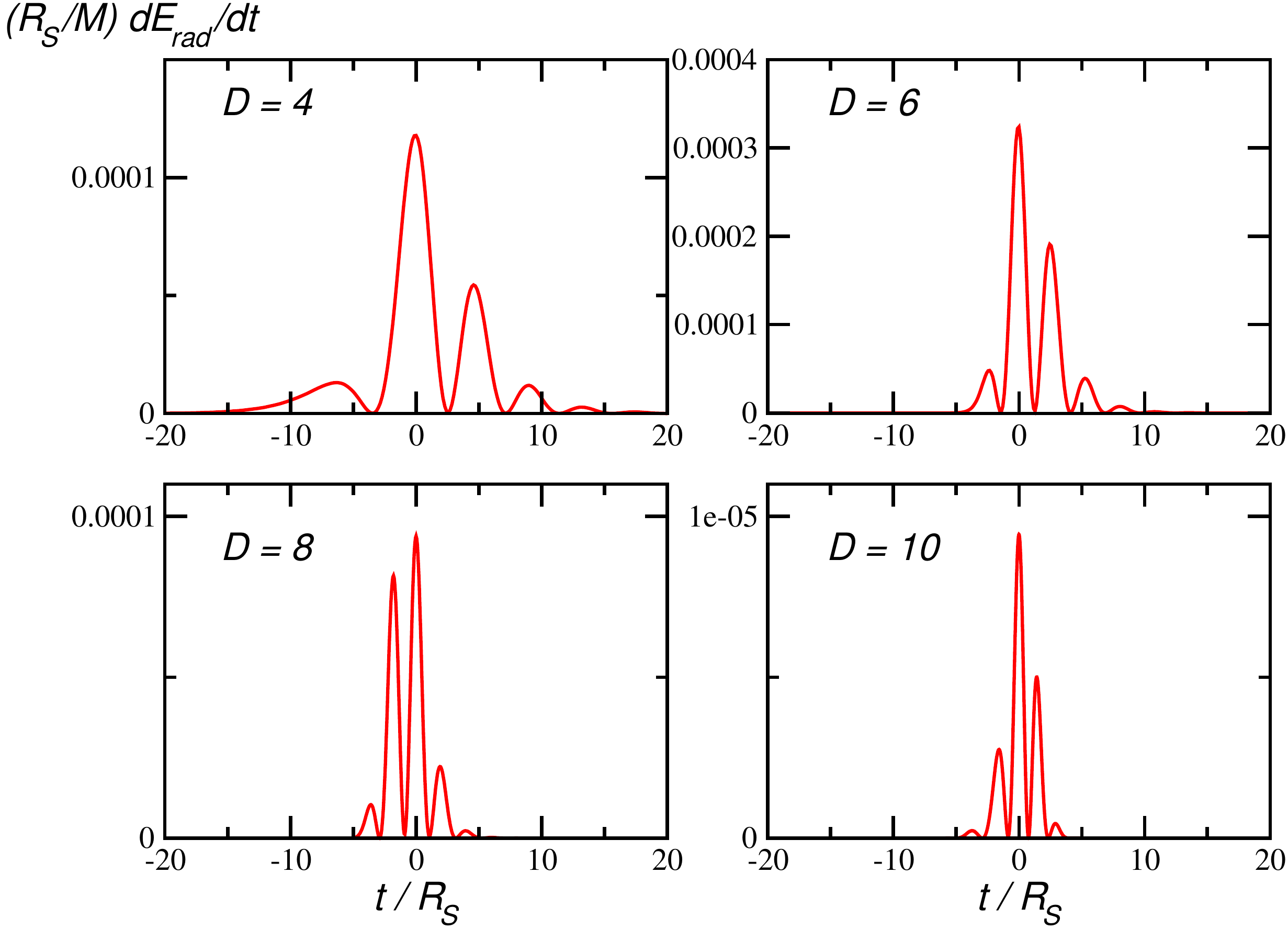}
  \caption{Normalized energy flux $(R_S/M) \dot{E}_{\rm rad}$ as a
    function of time for equal-mass collisions, with $t=0$ defined by
    the maximum in $\dot{E}_{\rm rad}$.  As $D$ increases, the burst
    of radiation becomes increasingly concentrated in time.}
  \label{fig:delta}
\end{figure}
\begin{figure}[ht]
\includegraphics[width=0.5\textwidth,clip]{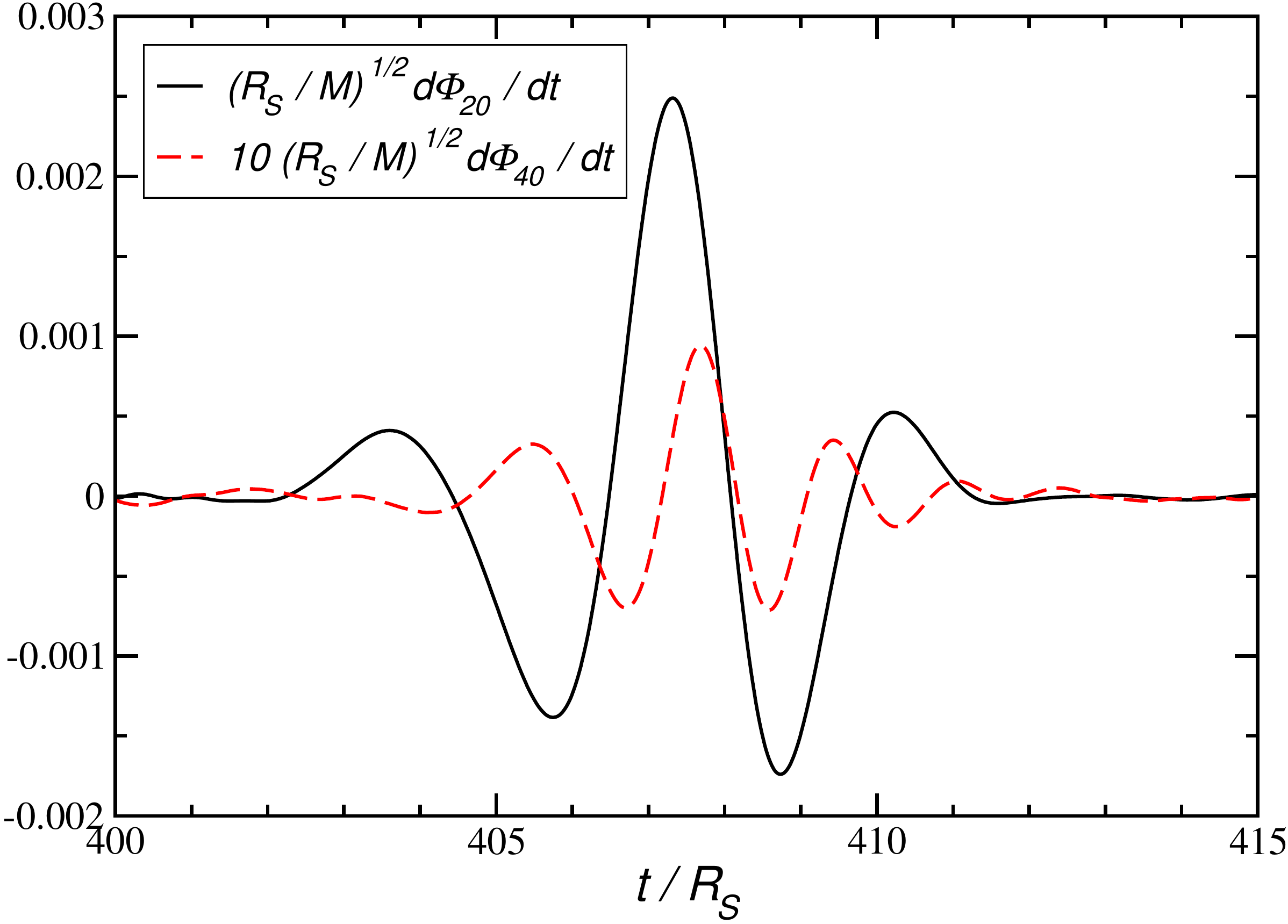}
\caption{The $l=2$ (solid black line) and $l=4$ (dashed red line)
  waveforms from the collision of two equal-mass BHs in $D=10$.}
\label{fig:equal_mass}
\end{figure}
\begin{figure}[tp]
\includegraphics[width=0.5\textwidth,clip]{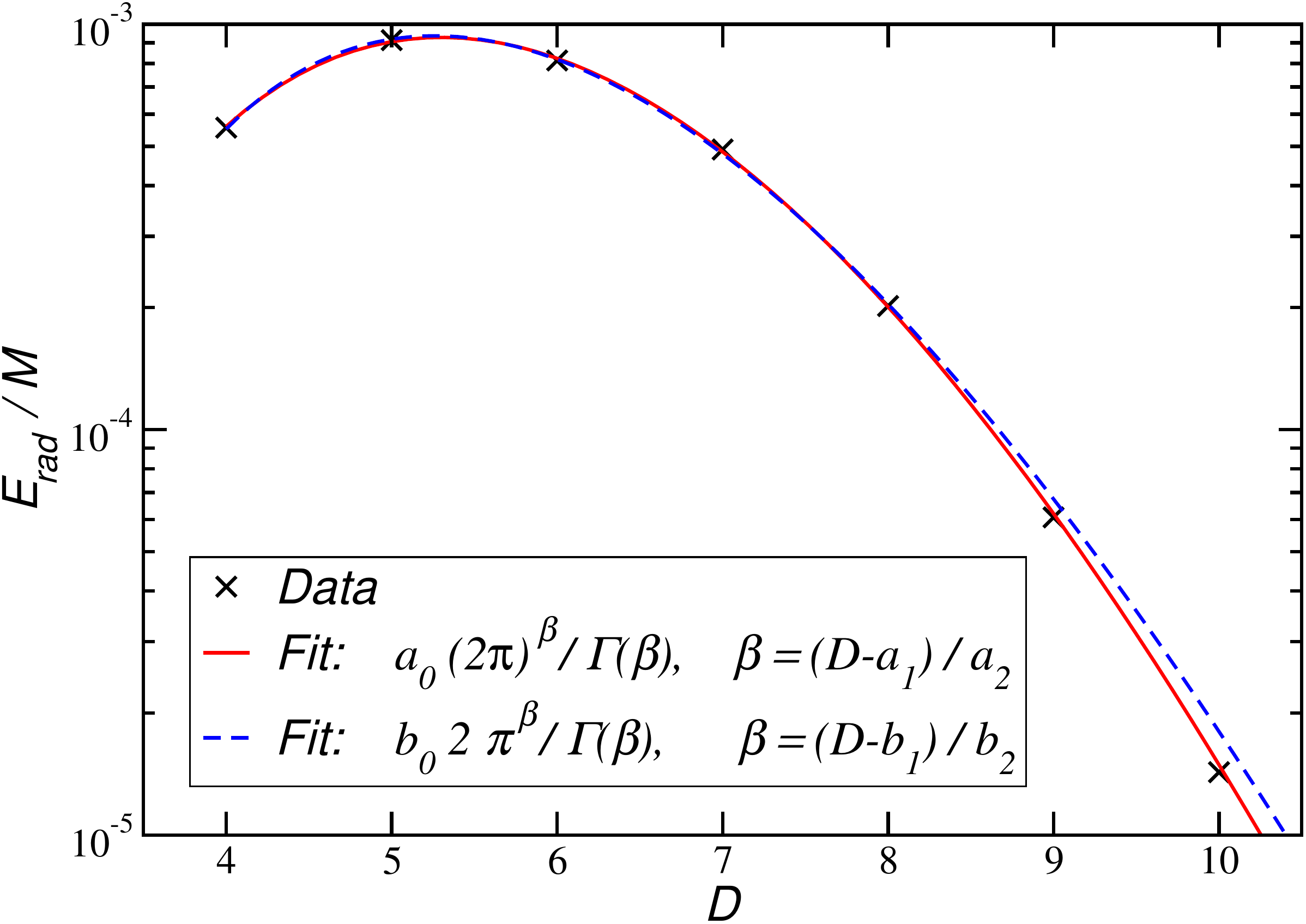}
\caption{Fractional energy $E_{\rm rad}/M$ radiated in GWs during
  collisions of equal-mass, nonspinning BHs starting from rest, in $D$
  spacetime dimensions. Crosses are numerical data points and the solid
  red line is the fit \eqref{q1fit}. The blue dashed line shows
  a fit obtained for the expression $b_0\,2\pi^{\beta}/\Gamma(\beta)$
  which resembles even more closely the functional form of the
  surface area $A_{D-2}=2\pi^{(D-1)/2}/\Gamma[(D-1)/2]$
  of the $D-2$ sphere, but does not match the data points as well.
  }
\label{fig:equal_mass_en}
\end{figure}
%

%%%%%%%%%%%%%%%%%%%%%%%%%%%%%%%%%%
\subsection{Equal-mass collisions}
\label{sec:results_EM}
%%%%%%%%%%%%%%%%%%%%%%%%%%%%%%%%%%
The collision of two equal-mass BHs has already been studied in
$D=4,\,5$~\cite{Witek:2010az}, and $D=6$~\cite{Witek:2014mha} spacetime
dimensions. We have verified those results, extending them to
$D=7,\,8\,,9\,,10$.
For illustration, in Fig.~\ref{fig:delta} we plot a normalized energy
flux $(R_S/M) \dot{E}$ for collisions in $D=4$, $6$, $8$ and $10$
spacetime dimensions. As $D$ increases, the burst of radiation becomes
increasingly concentrated in time.  This concentration suggests
that the burst may approach a $\delta$
distribution in the large-$D$ limit; it would be interesting to see if
this is borne out in the large-$D$ limit formalism of
\cite{Emparan:2013moa,Emparan:2015hwa,Emparan:2015gva,Emparan:2014cia}.

For further illustration, in 
Fig.~\ref{fig:equal_mass} we plot the Kodama-Ishibashi waveform
$\dot{\Phi}_{l0}$
\cite{Kodama:2003jz,Ishibashi:2011ws,Witek:2010xi,Witek:2014mha}
for $D=10$; the qualitative features of the signal are the same for
all other $D$.  The waveform consists of a precursor part with small
amplitude when the two BHs are widely separated, followed by a smooth
merger phase connecting to ringdown. A perturbative calculation, 
using direct integration techniques, yields
the following two modes for gravitational-type scalar perturbations:
$\omega R_S=1.2346- 0.9329i$ and $\omega R_S=2.4564-0.9879i$. These
are the decoupling (or saturating) and nondecoupling (or nonsaturating) modes in the language
of Refs.~\cite{Emparan:2014aba,Emparan:2015rva} (Ref.~\cite{Dias:2014eua}). We find agreement to the level of« $\sim 0.1\%$ or better
with Ref.~\cite{Dias:2014eua} and very good agreement with the analytical, large-$D$ estimates of Ref.~\cite{Emparan:2014aba}.
A one-mode fit of numerical waveforms yields very poor agreement with
any of the frequencies above.  However, a two-mode fit yields the
following two frequencies: $\omega R_S=2.48 -0.94i,\, 1.22- 0.91i$.
Given the errors in numerical simulations, this is a reasonable level
of agreement with linearized predictions, and it indicates that both
modes are excited to comparable amplitudes for this particular
simulation.

When plotted as a function of the number $D$ of dimensions
(Fig.~\ref{fig:equal_mass_en}), the fraction of center-of-mass energy
radiated in GWs by equal-mass head-on collisions reaches a maximum
$E_{\rm rad}/M\sim 9.1 \times 10^{-4}$ for $D=5$.  Beyond this value,
we find the total radiation output to rapidly decrease as a function
of $D$.  This suppression is consistent with the fact that the
spacetime is nearly flat outside the horizon: in fact, the
gravitational potential $(R_S/r)^{D-3}$ vanishes exponentially with
$D$~\cite{Emparan:2013moa}.
Another intuitive explanation for this rapid decay is that, as $D$
increases, the energy is radiated almost instantaneously
(cf. Fig.~\ref{fig:delta}): spacetime is flat except extremely near the horizons,
and bremsstrahlung radiation is suppressed. These features have also been seen in zero-frequency limit calculations~\cite{Cardoso:2002pa}. Thus, at large $D$, radiation is emitted in a burst precisely when the BHs collide, but this is also the instant where one would expect common horizon formation, and consequent absorption of a sizable fraction of this energy. This is, of course, a very loose description, unable to give us a quantitative estimate.
The results in Fig.~\ref{fig:equal_mass_en} are (perhaps surprisingly)
well described by the following simple analytic expression,
\begin{eqnarray}
  &&\frac{E_{\rm rad}^{q=1}}{M}=a_0
        \frac{\left(2\pi\right)^{\beta}}{\Gamma[\beta]}\,,
     \quad
        \beta=\frac{D-a_1}{a_2}\,,
        \label{q1fit}
\end{eqnarray}
where $a_0=1.7288\times 10^{-6}$, $a_1 = 1.5771$, $a_2 = 0.5497$.
This fit reproduces our numerical results to
within $\sim 1\%$ for all $D=4,\dots,10$. It is tempting to relate
this expression to the area
$A_{D-2} = \frac{2\pi^{(D-1)/2}} {\Gamma(\frac{D-1}{2})}$ 
of a $(D-2)$-dimensional unit sphere, but we do
not see an evident connection as the numerical factors do not match
exactly\footnote{The expression $b_0\,2\pi^{\beta}/\Gamma(\beta)$
resembles even more closely that of the surface area $A_{D-2}$,
but yields a less accurate fit to the data points
(cf.~Fig.~\ref{fig:equal_mass_en}).
It also does not establish 
a satisfactory relation between $A_{D-2}$
and the numerical parameters appearing in the fit for
$\beta$, now given by $\beta=(D-2.4772)/0.7671$.}.

The results for the radiated energy are in stark contrast to the
predictions one would get by applying the PP results of Table
\ref{tab:summaryPP} to the equal-mass case $q=1$, where, instead of a
strong suppression of $E_{\rm rad}$ at large $D$, we see a mild
increase in the radiative efficiency.  While the PP approximation is
by construction not expected to capture the equal-mass limit with high
precision, it is valuable to understand the origin of this qualitative
discrepancy.  A tantalizing suggestion in this context was made by
Emparan and collaborators~\cite{Emparan:2013moa}, who pointed out that --
for large $D$ -- BH spacetimes contain two scales ${\cal L}$ of
interest for BH physics. One scale can be parametrized by the areal
radius ${\cal L}\sim R_S$ of the horizon.  The other scale, absent at
low $D$, is related to the strong localization of the gravitational
potential close to the horizon: ${\cal L}\sim R_S/D$.  For equal-mass
collisions the excitation of the latter modes (and the radiation
output) are strongly suppressed at large $D$~\cite{Emparan:2013moa}.
However, dynamical processes are very sensitive to the dominant scale
in higher dimensions~\cite{Cardoso:2002pa,Emparan:2013moa}.  In the
next section, we explore in more detail unequal-mass collisions and
indeed find that these collisions can trigger the excitation of
smaller-scale modes even at the low energies considered in our
simulations.
%
%%%%%%%%%%%%%%%%%%%%%%%%%%%%%%%%%%%%%%%%%%%%%%%%%%%%%%%%%%%%%%%%%
\subsection{Unequal-mass collisions and the point-particle limit}
%%%%%%%%%%%%%%%%%%%%%%%%%%%%%%%%%%%%%%%%%%%%%%%%%%%%%%%%%%%%%%%%%
%
\begin{figure*}[tb]
  \begin{tabular}{cc}
  \includegraphics[width=0.5\textwidth,clip]{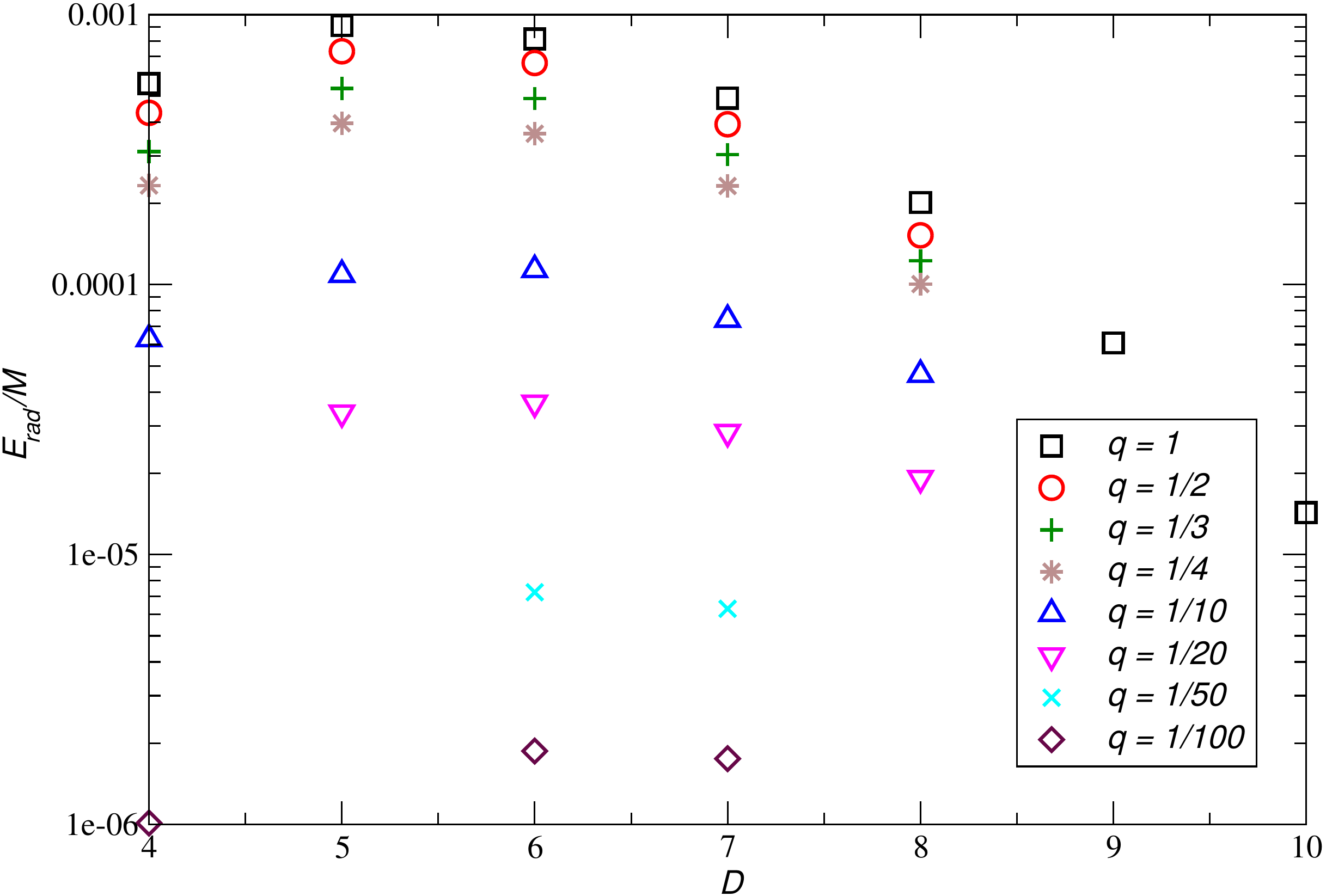}&
  \includegraphics[width=0.5\textwidth,clip]{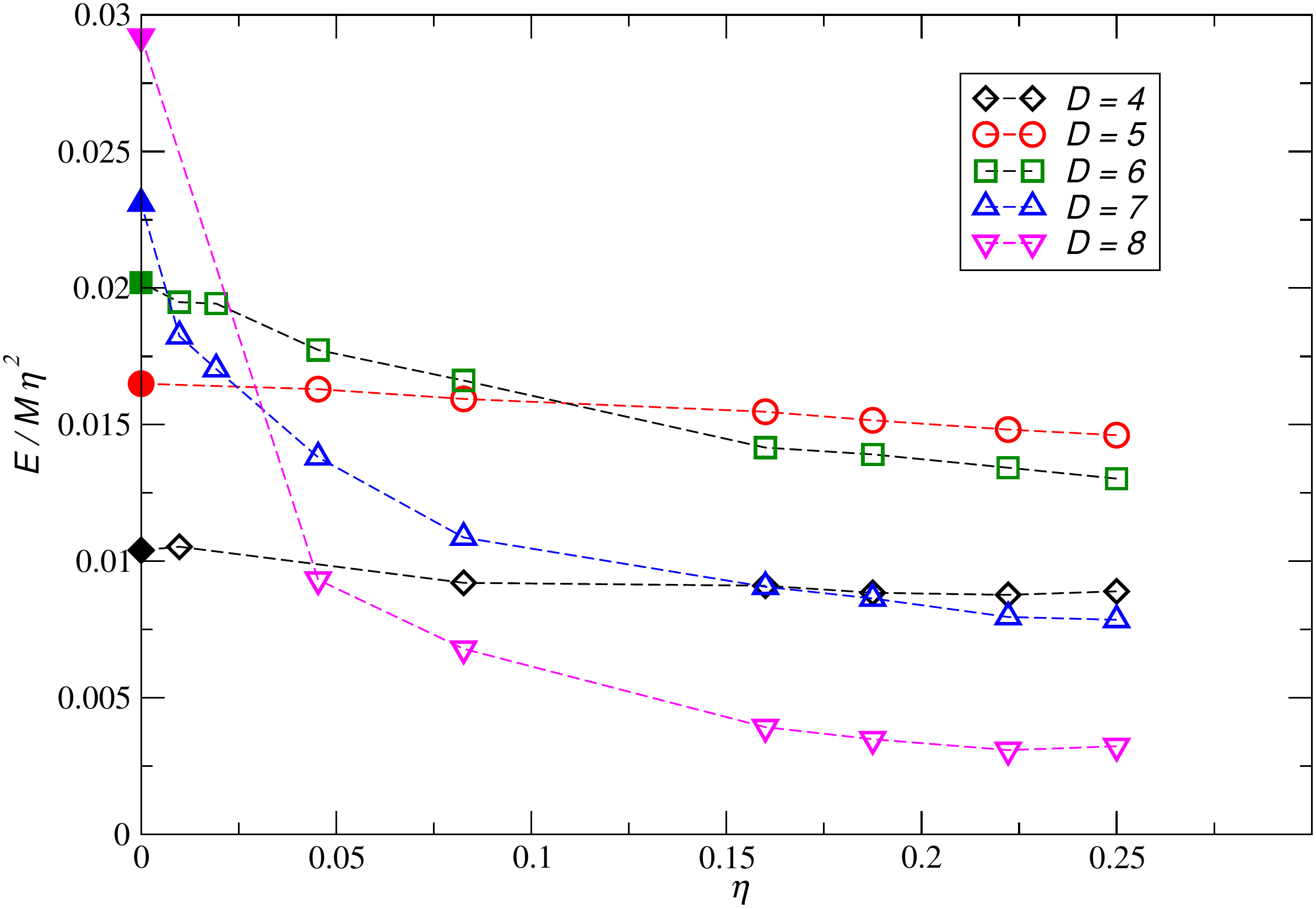}
  \end{tabular}
  \caption{Left panel: fractional energy $E_{\rm rad}/M_{\rm ADM}$
    radiated in GWs in collisions of nonspinning BHs starting from
    rest with mass ratio $q$ in $D$ spacetime dimensions. Right panel:
    same data as in the left panel, but rescaled by $\eta^2$ [i.e. we
    plot $E_{\rm rad}/(M\eta^2)$] in order to facilitate the
    comparison with PP calculations of the radiated energy, which are
    shown as filled symbols at $\eta=0$. }
  \label{fig:unequal_E}
\end{figure*}
The stark contrast between the PP results summarized in
Table~\ref{tab:summaryPP} and the numerical relativity calculations of
the previous section strongly points towards a qualitatively
different behavior of the radiated energy as a function of $D$
for comparable-mass binaries (where $E_{\rm rad}$ rapidly drops beyond
$D=6$) as compared with the high mass-ratio regime (where $E_{\rm rad}$
mildly increases with $D$). The question we are now facing is:
does the difference in the behavior arise from the dominance of
different physical mechanisms in the respective regions of the
parameter space, and where
does the crossover from one regime to the other occur?
To shed light on this
issue, we have performed collisions of unequal-mass, nonspinning BHs
focusing on the range $q=1,\dots,1/100$ and $D=4,\dots,8$.  The GW
energy and linear momentum radiated in these collisions are summarized
in Figs.~\ref{fig:unequal_E}-\ref{fig:Pofq}.

By analyzing the waveforms for the most extreme mass ratios we find
good agreement between the ringdown stage and estimates from
linearized perturbations. However, our results indicate that only the high-frequency
modes (the ``nonsaturating'' modes) are excited. Since these modes probe the small scales
presumably excited by the smaller BH~\cite{Emparan:2015hwa,Emparan:2014cia}, it is reassuring to find high-frequency excitations.

Figure~\ref{fig:unequal_E} shows the fractional center-of-mass energy released as GWs when two BHs collide,
with and without normalization by (the square of) the kinematic, symmetric mass ratio parameter $\eta$.
Note that $\eta$ is directly connected to
the reduced mass of the system and is known to yield a very good
rescaling of all quantities in four-dimensional spacetimes (see for
instance Refs.~\cite{Berti:2007fi,LeTiec:2011bk,Tiec:2014lba}).  For
low $D$ (in particular for $D=4,\,5$) the total radiated energy
$E_{\rm rad}/(M\eta^2)$ is weakly dependent on $\eta$. At small mass
ratios $q$, or equivalently at small $\eta$, our results smoothly
approach the PP limit of Table~\ref{tab:summaryPP} (shown in
Fig.~\ref{fig:unequal_E} as filled data points at $\eta=0$).

For $q\lesssim 1$ and sufficiently large $D$, the radiated energy decreases
monotonically with $D$ (left panel of Fig.~\ref{fig:unequal_E}). This
behavior would clearly contradict the PP results if it held for
arbitrarily small mass ratio.  In fact, at small mass ratios the
behavior of the radiated energy changes.  The maximum of the radiated
energy as a function of $D$ shifts from $D=5$ to $D=6$ between $q=1/4$
and $q=1/10$. Results for even smaller $q$ indicate a further shift
towards $D=7$, and possibly yet higher $D$ as we approach the PP
limit. Furthermore, we see from the right panel of
Fig.~\ref{fig:unequal_E} that $E_{\rm rad}/(M\eta^2)$ shows a steep
increase for very small $\eta$ and large $D$. This behavior supports our
interpretation that new scales are being probed.  If this is indeed
the correct interpretation, and if the new scale is of order $R_S/D$,
one can estimate the mass ratio at which these new scales are excited.
By using Eq.~(\ref{eq:RSMADM}), and recalling that $M_2/M_1=q$, we get
the scaling $\left(r_2/R_S\right)^{D-3}=q$, with $r_2$ the scale of
the small BH and $R_S$ the scale of the large BH in terms of
coordinate quantities.  If we equate the ``small scale'' $R_S/D$ to
the size $r_2$ of the second colliding object we find the threshold
mass ratio
\be
q\sim D^{3-D}\,.
\ee
\begin{table}[t!]
\centering
\begin{tabular}{c|ccccc}
\hline
\hline
$D$                         &  4          & 5   & 6      & 7      & 8\\
\hline
$-10^2b_1$                &  0.54    & 0.95    & 2.82    & 3.63 & $3.58$\\
$b_2$                     &  0.72    & 1.18    & 0.83    & 0.44 & $0.19$\\
\hline
\hline
\end{tabular}    
\caption{\footnotesize Fitting coefficients of Eq.~\eqref{fit_PP},
  describing the $\eta$ dependence of the total radiated energy.}
\label{tab:summaryPP_coeff}
\end{table}
It seems sensible to understand the mass ratio dependence by {\it
  fixing} the PP limit to be that of Table~\ref{tab:summaryPP}.  In
other words, we fit our results to the expression
\be
\frac{E_{\rm rad}}{M\eta^2}=b_0+b_1\eta^{b_2}\,,\label{fit_PP}
\ee
where $b_0$ are the PP values listed in Table~\ref{tab:summaryPP}. The
exponents $b_2$ obtained by fitting our data are listed in
Table~\ref{tab:summaryPP_coeff}. These numbers are consistent with the
behavior shown in Fig.~\ref{fig:unequal_E}: the dependence of the
total radiated energy on $\eta$ is more complex for large $D$. In
particular, at large $D$ the expansion of $E_{\rm rad}$ in powers of
$\eta$ converges more slowly, and the convergence of the PP results (a
leading-order expansion in mass ratio) is poor in the small-$\eta$
regime.  It would be interesting to find an analytical prediction for
the coefficient $b_2$.

%%%%%%%%%%%%%%%%%%%%%%%%%%%%%%%%%%%%%%%%%%%%%%%%%%%%%%%%%%%%%%%%%
\subsection{Kicks}
%%%%%%%%%%%%%%%%%%%%%%%%%%%%%%%%%%%%%%%%%%%%%%%%%%%%%%%%%%%%%%%%%
%
\begin{figure}[tb]
  \includegraphics[width=0.5\textwidth,clip]{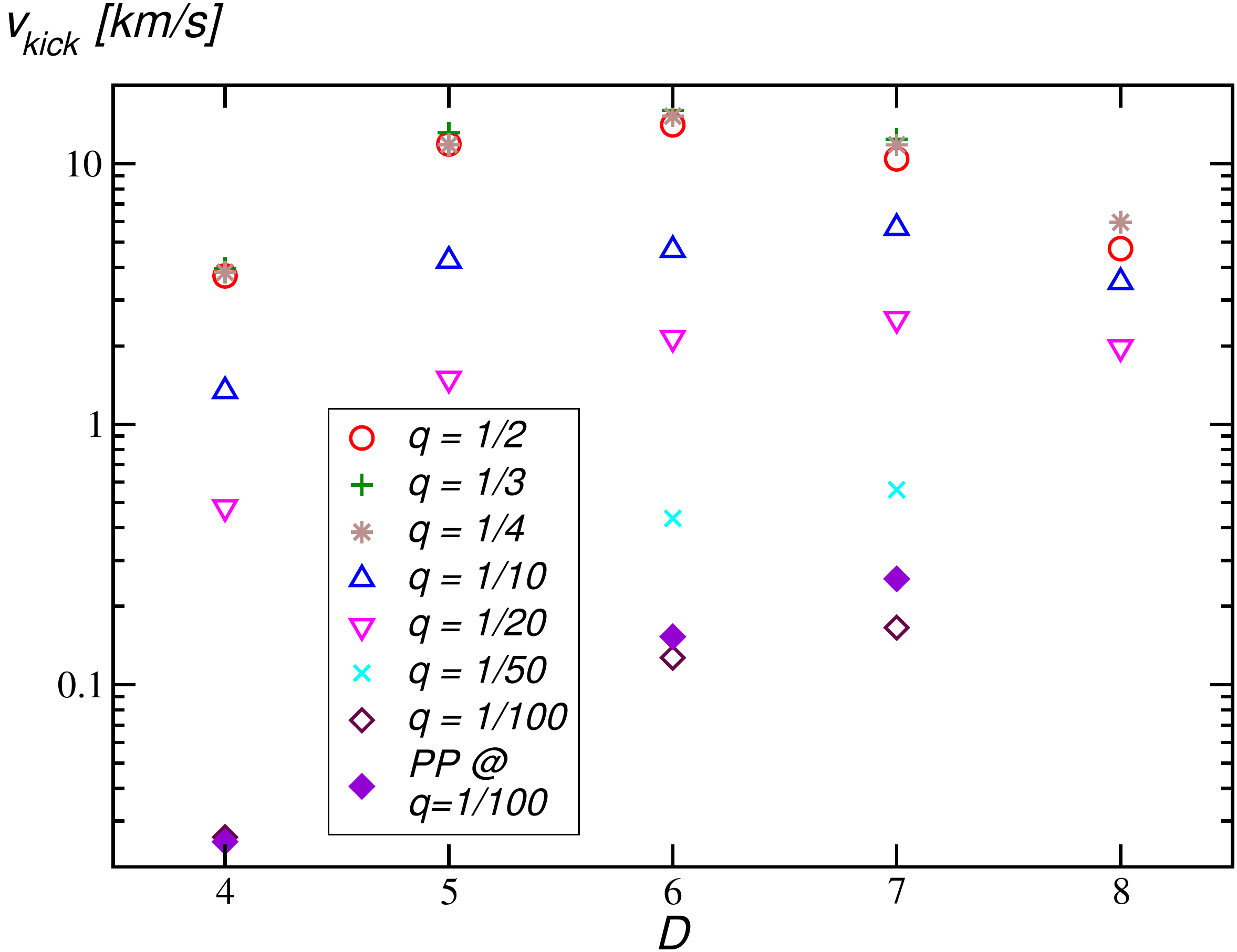}
  \caption{Recoil due to asymmetric emission of GWs in the collision
    of nonspinning BHs starting from rest with mass ratio $q$ in $D$
    spacetime dimensions. Note that the agreement with PP predictions
    in the small-$q$ limit is very good for $D=4$, and degrades for
    higher $D$.}
  \label{fig:PofD}
\end{figure}
\begin{figure}[tb]
  \includegraphics[width=0.5\textwidth,clip]{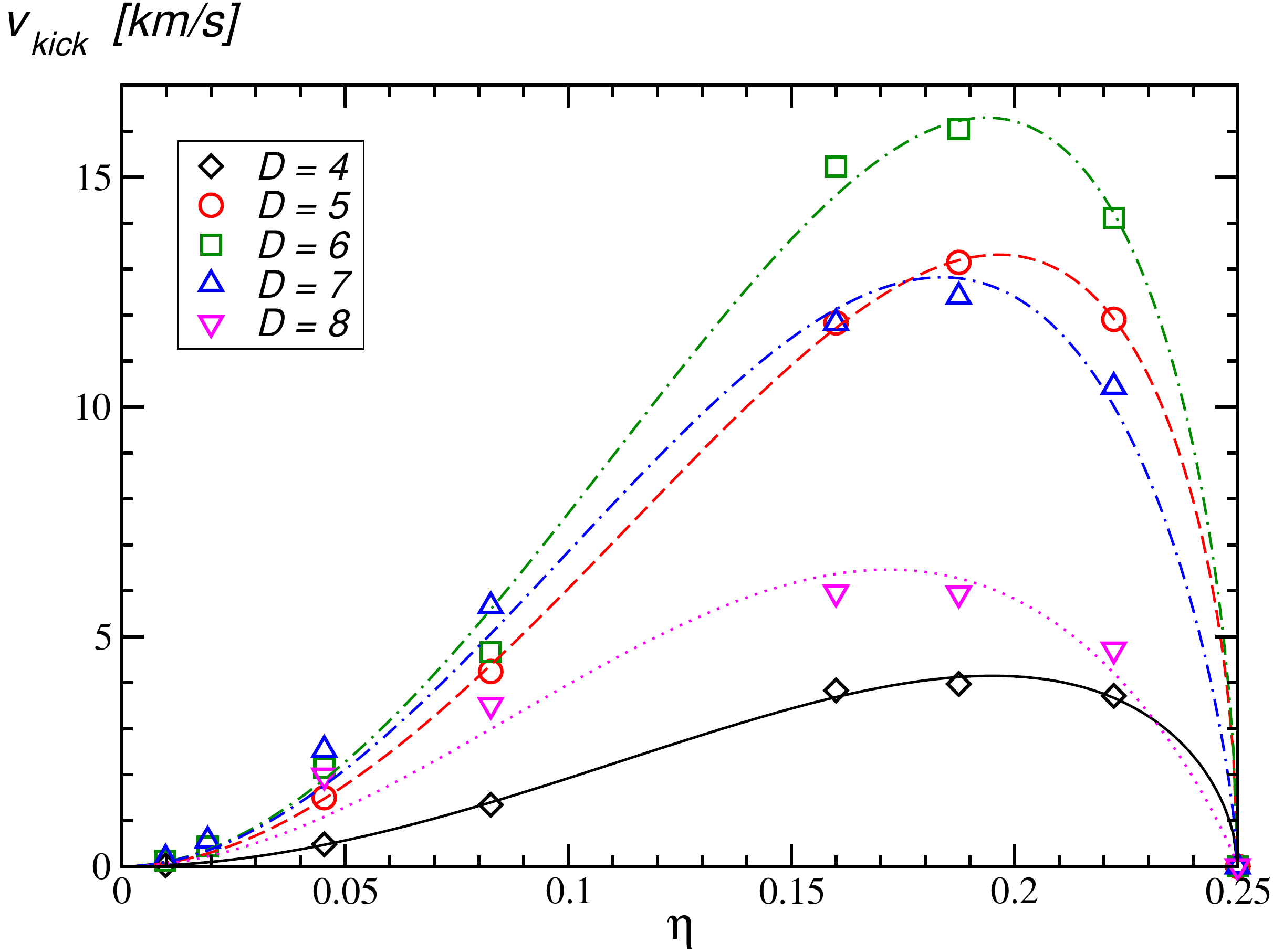}
  \caption{As Fig.~\ref{fig:PofD} but here symbols denote the kick for
    fixed $D$ as a function of the symmetric mass ratio $\eta$. The
    lines are the simple two-parameter fit of Eq.~\eqref{kickfit2}.}
  \label{fig:Pofq}
\end{figure}
In Fig.~\ref{fig:PofD}, we show the gravitational recoil (or ``kick'')
velocity of the postmerger BH as a function of $D$ for fixed values
of the mass ratios $q$.  As in the case of the radiated energy (left
panel of Fig.~\ref{fig:unequal_E}), we observe a shift in the maximum
kick towards higher $D$ as the mass ratio decreases. In particular,
the maximum shifts from $D=6$ to $D=7$ as we change $q$ from $1/4$ to
$1/10$. In Fig.~\ref{fig:Pofq} we show the same
results, but now plotting the kick for fixed $D$ as a function of the
symmetric mass ratio $\eta$.

The data in Figs.~\ref{fig:PofD} and \ref{fig:Pofq} are in good
agreement with PP recoil
calculations~\cite{Nakamura:1983hk,Berti:2010ce,Berti:2010gx}: for
example, in $D=4$ the PP calculation yields
$P_{\rm rad}/M=8.33\times 10^{-4} q^2$, or
$v_{\rm kick}=250 q^2$~km/s. This is in percent-level agreement with
the $D=4$, $\eta=0.01$ simulation, for which we get
$v_{\rm kick}=0.026$~km/s (for such small mass ratios, of course,
$q\simeq \eta=0.01$).  As $D$ increases, the PP prediction becomes
less accurate: the relative error is $4\%$ in $D=4$, $21\%$ in $D=6$
and $54\%$ in $D=7$.  This is consistent with the trend observed for
the radiated energy and with physical expectations: according to
Eq.~(\ref{eq:RSMADM}), for a fixed $q$ the less massive black hole
appears less and less like a PP. It is also possible that some of this
disagreement comes form the larger errors in the high-$D$, small-mass
ratio simulations.

Following previous work on unequal mass collisions in $D=4$ dimensions
\cite{Sperhake:2011ik}
  we first tried to fit the data using the following mass
ratio dependence (see e.g. the classic work by Fitchett and Detweiler
\cite{Fitchett:1984qn}):
\begin{equation}
  v^{(1)}_{\rm kick} %= v_D \frac{q^2(q-1)}{(1+q)^5}
  =v_D \eta^2\sqrt{1-4\eta}\,,
  \label{kickfit1}
\end{equation}
where the superscript $(1)$ means that this is a one-parameter
fit. According to this simple formula, the maximum recoil occurs when
$\eta=0.2$ ($q\simeq 0.38$) for all $D$. Note that for $\eta=0.2$ we
get $v^{(1)}_{\rm kick, max}\simeq 0.018 v_D$, so the parameter $v_D$
is related to the maximum kick by a simple proportionality relation.

However, our previous considerations suggest that the mass ratio
dependence of the radiated energy and of the recoil velocity should
vary with $D$. As a simple way to investigate this $D$ dependence we
used a two-parameter fitting function:
\begin{equation}
  v^{(2)}_{\rm kick} = \tilde v_D \eta^2 (1-4\eta)^{c_D}\,.
  \label{kickfit2}
\end{equation}
Assuming this dependence, the maximum kick $v^{(2)}_{\rm kick, max}$
will correspond to a $D$-dependent $\eta_{\rm max}$ that can be
obtained by fitting the data.

The fitting coefficients and maximum kicks obtained with these two
expressions are listed in Table~\ref{tab:kicks}. Note that the $D$
dependence of $\eta_{\rm max}$ is very mild for all but the largest
$D$ simulations. More accurate simulations may be needed to resolve
the issue of the $D$-dependence of $\eta_{\rm max}$ and of the maximum
kick velocity. However, the following conclusion is quite independent
of the assumed functional dependence: the maximum kick is
$\sim 16.3$~km/s, and it is achieved for $D=6$ and
$\eta_{\rm max}\simeq 0.2$.
\begin{table}[t!]
\centering
\begin{tabular}{c|ccccc}
\hline
\hline
$D$                         &  4          & 5      & 6      & 7      & 8\\
\hline 
$v_D~[{\rm km/s}]$          &  232.9    & 746.9  & 915.2 & 714.8 & 349.0 \\
$v^{(1)}_{\rm kick, max}~[{\rm km/s}]$ &  4.166 & 13.361 & 16.372 & 12.787 &  6.244 \\
 \hline 
$\tilde v_D~[{\rm km/s}]$  &  255.8    & 798.4 & 1034 & 989.9 & 630.7\\
$c_D$                         &  0.5629    & 0.5445 & 0.5821 & 0.7214 & 0.9110\\
$\eta_{\rm max}$          &  0.1951 & 0.1965 & 0.1936 & 0.1837 & 0.1718\\
$v^{(2)}_{\rm kick, max}$[km/s]  & 4.148 & 13.314 & 16.297 & 12.822 & 6.457 \\
\hline
\hline
\end{tabular}
\caption{\footnotesize Fitting coefficients of Eqs.~\eqref{kickfit1}
  and \eqref{kickfit2}, describing the $\eta$ dependence of the kick velocity.}
\label{tab:kicks}
\end{table}
%

%%%%%%%%%%%%%%%%%%%%%%%%
\section{Conclusions}
\label{sec:conclusions}
%%%%%%%%%%%%%%%%%%%%%%%
We have numerically simulated head-on collisions of black holes in
$D=4,\,\ldots,\,10$ dimensions, extracted the GW signal and computed
the energy and linear momentum radiated in the collisions. Starting
with the equal-mass case, we find values for the radiated energy in
agreement with previously published results for $D=5$ and $D=6$
dimensions. The radiated energy, measured in units of the ADM mass
$M$, is maximal in $D=5$, where $E_{\rm rad}/M =
9.1\times10^{-4}$. For larger $D$ we observe a strong reduction in
the radiated energy: the fit
$E_{\rm rad}/M = (2\pi)^{\beta}/ \Gamma(\beta),~
\beta=(D-1.5771)/(0.5497)$ models our results to within $1\,\%$ for
all $D$ simulated. This functional dependence closely resembles that
of the surface area
$\mathcal{A}_{D-2}=2\pi^{(D-1)/2}/\Gamma[(D-1)/2]$, but the
discrepancy in the numerical parameters in the argument suggests a
more complicated relation between the two quantities.

The numerical results for the equal-mass case differ strikingly
from those obtained in the PP approximation, which predicts a mild increase
of $E_{\rm rad}/(q^2M)$ with $D$ when a small BH of mass
$qM_1,~q\ll 1$ falls into a BH of mass $M_1$. We reconcile these
seemingly different predictions by numerically simulating a wider set
of BH collisions with mass ratios ranging from $q=1$ to $q=1/100$ in
up to $D=8$ dimensions. In the right panel of Fig.~\ref{fig:unequal_E}
we observe that the (symmetric mass ratio-normalized) energy
$E_{\rm rad}/(M\eta^2)$ increases in the PP limit $q\rightarrow 0$.
This increase becomes particularly steep for $D=7$ and $D=8$, and the
numerical data extrapolated to $q=0$ are in good agreement with the PP
predictions.

These findings can be understood by invoking the presence of multiple
length scales in the large-$D$ limit, as identified in
\cite{Emparan:2013moa}: Additionally to the length scale $R_S$ of the
Schwarzschild horizon, the large-$D$ limit reveals a shorter scale
$R_S/D$ for the spatial variation of potential terms in the equations
governing BH perturbations. It is natural then to assume that these
shorter length scales will be excited with much higher efficiency by a
small object falling into a BH, while they are largely insensitive to
the collision of two objects of size $R_S$. The parameter
regime in between these two extremes, on the other hand, is characterized
by excitations of comparable magnitude on both length scales.

Our intuitive interpretation is strengthened by the analysis of the
quasinormal mode frequencies:
for $q=1$ (and large $D$) the ringdown exhibits comparable
contributions from two frequencies, corresponding to the
``saturating'' and ``unsaturating'' modes in the language of
\cite{Dias:2014eua}, while the ringdown is dominated by the
unsaturating modes for $q\ll 1$.
For large $D$, the emission of gravitational waves therefore appears to
be sensitive to the properties of the two BHs.
It is interesting to contrast this observation
with the corresponding {\em insensitivity} of the collision dynamics
in high-energy collisions
in $D=4$ \cite{Sperhake:2012me,Sperhake:2015siy}. This contrast
naturally raises the question which effect dominates in high-energy,
large-$D$ collisions: sensitivity to structure due to large $D$ or
universality due to high energy?

With regard to the large-$D$ limit, we notice a further connection
in the shape of
the energy flux as a function of time. In units of the Schwarzschild
horizon associated with the ADM mass of the spacetime, the flux
becomes increasingly peaked in higher $D$ and it appears to approach
the shape of a $\delta$ distribution, which is what one would
intuitively expect in the large-$D$ limit, where the spacetime
exterior to a BH approaches Minkowski.

Finally, we analyze the gravitational recoil resulting from the
asymmetric emission of GWs in unequal-mass collisions. We find the
data to be well fitted by Fitchett's \cite{Fitchett:1984qn} formula
commonly applied to the four-dimensional case, but we also observe a
mild indication that the mass ratio maximizing the recoil varies with
$D$ at large $D$. The maximum kick due to gravitational recoil
($v_{\rm kick,max}\sim 16.3$~km/s) is achieved for $D=6$, and for a
symmetric mass ratio $\eta=\eta_{\rm max}\simeq 0.2$ ($q\simeq 0.4$).
When regarding both energy or linear momentum as a function of $D$ at
fixed mass ratio $q$, we observe a shift in the maximum towards higher
$D$ as we move from the equal-mass case $q=1$ to the PP limit
$q\ll 1$.  This observation further confirms one of our main
conclusions: the PP limit provides exquisitely accurate predictions
for small mass ratios, but it must be taken with a grain of salt when
extrapolated to the comparable-mass regime in higher dimensions.

%%%%%%%%%%%%%%%%%%%%%%%
\begin{acknowledgments}
We are grateful to Roberto Emparan for numerous suggestions and for sharing with us some numerical results.
We thank Pau Figueras, Markus Kunesch, Chris Moore, Saran Tunyasuvunakool, Helvi Witek
and Miguel Zilh{\~a}o for very fruitful discussions on this topic.
V.C. is indebted to Kinki University in Osaka for hospitality while the last stages of this work were being completed.
U. S. and V. C. acknowledge financial support provided under the European Union's H2020 ERC Consolidator Grant ``Matter and strong-field gravity: New frontiers in Einstein's theory'' Grant Agreement No. MaGRaTh--646597. Research at Perimeter Institute is supported by the Government of Canada through Industry Canada and by the Province of Ontario through the Ministry of Economic Development $\&$
Innovation.
E.~B. was supported by NSF Grants Nos.~PHY-1607130 and AST-1716715, and
by FCT Contract IF/00797/2014/CP1214/CT0012 under the IF2014
Programme.
This work has received funding from the European Union's Horizon
2020 research and innovation programme under the Marie Sk\l odowska-Curie
Grant Agreement No.~690904,
the COST Action Grant No.~CA16104,
from STFC Consolidator Grant No. ST/L000636/1,
the SDSC Comet, PSC-Bridges and TACC Stampede clusters through NSF-XSEDE Award
No.~PHY-090003,
the Cambridge High Performance Computing Service Supercomputer Darwin using Strategic Research
Infrastructure Funding from the HEFCE and the STFC, and DiRAC's Cosmos
Shared Memory system through BIS Grant No.~ST/J005673/1 and STFC Grant
Nos.~ST/H008586/1, ST/K00333X/1.
W.G.C. is supported by a STFC studentship.
We acknowledge PRACE for awarding us access to
MareNostrum at Barcelona Supercomputing Center (BSC), Spain
under Grant No.~2016163948.
\end{acknowledgments}
%%%%%%%%%%%%%%%%%%%%%%%%%%%%%%%%%%%%%%%%%%%%%%%%%%%%%%%%%%%%%%%%

%%%%%%%%%%%%%%%%%%%%%
%\section*{References}
%%%%%%%%%%%%%%%%%%%%%
%\bibliography{newuli}

\begin{thebibliography}{60}%
\makeatletter
\providecommand \@ifxundefined [1]{%
 \@ifx{#1\undefined}
}%
\providecommand \@ifnum [1]{%
 \ifnum #1\expandafter \@firstoftwo
 \else \expandafter \@secondoftwo
 \fi
}%
\providecommand \@ifx [1]{%
 \ifx #1\expandafter \@firstoftwo
 \else \expandafter \@secondoftwo
 \fi
}%
\providecommand \natexlab [1]{#1}%
\providecommand \enquote  [1]{``#1''}%
\providecommand \bibnamefont  [1]{#1}%
\providecommand \bibfnamefont [1]{#1}%
\providecommand \citenamefont [1]{#1}%
\providecommand \href@noop [0]{\@secondoftwo}%
\providecommand \href [0]{\begingroup \@sanitize@url \@href}%
\providecommand \@href[1]{\@@startlink{#1}\@@href}%
\providecommand \@@href[1]{\endgroup#1\@@endlink}%
\providecommand \@sanitize@url [0]{\catcode `\\12\catcode `\$12\catcode
  `\&12\catcode `\#12\catcode `\^12\catcode `\_12\catcode `\%12\relax}%
\providecommand \@@startlink[1]{}%
\providecommand \@@endlink[0]{}%
\providecommand \url  [0]{\begingroup\@sanitize@url \@url }%
\providecommand \@url [1]{\endgroup\@href {#1}{\urlprefix }}%
\providecommand \urlprefix  [0]{URL }%
\providecommand \Eprint [0]{\href }%
\providecommand \doibase [0]{http://dx.doi.org/}%
\providecommand \selectlanguage [0]{\@gobble}%
\providecommand \bibinfo  [0]{\@secondoftwo}%
\providecommand \bibfield  [0]{\@secondoftwo}%
\providecommand \translation [1]{[#1]}%
\providecommand \BibitemOpen [0]{}%
\providecommand \bibitemStop [0]{}%
\providecommand \bibitemNoStop [0]{.\EOS\space}%
\providecommand \EOS [0]{\spacefactor3000\relax}%
\providecommand \BibitemShut  [1]{\csname bibitem#1\endcsname}%
\let\auto@bib@innerbib\@empty
%</preamble>
\bibitem [{\citenamefont {Antoniadis}(1990)}]{Antoniadis:1990ew}%
  \BibitemOpen
  \bibfield  {author} {\bibinfo {author} {\bibfnamefont {I.}~\bibnamefont
  {Antoniadis}},\ }\href {\doibase 10.1016/0370-2693(90)90617-F} {\bibfield
  {journal} {\bibinfo  {journal} {Phys. Lett. B}\ }\textbf {\bibinfo {volume}
  {246}},\ \bibinfo {pages} {377} (\bibinfo {year} {1990})}\BibitemShut
  {NoStop}%
\bibitem [{\citenamefont {Arkani-Hamed}\ \emph {et~al.}(1998)\citenamefont
  {Arkani-Hamed}, \citenamefont {Dimopoulos},\ and\ \citenamefont
  {Dvali}}]{ArkaniHamed:1998rs}%
  \BibitemOpen
  \bibfield  {author} {\bibinfo {author} {\bibfnamefont {N.}~\bibnamefont
  {Arkani-Hamed}}, \bibinfo {author} {\bibfnamefont {S.}~\bibnamefont
  {Dimopoulos}}, \ and\ \bibinfo {author} {\bibfnamefont {G.~R.}\ \bibnamefont
  {Dvali}},\ }\href {\doibase 10.1016/S0370-2693(98)00466-3} {\bibfield
  {journal} {\bibinfo  {journal} {Phys. Lett. B}\ }\textbf {\bibinfo {volume}
  {429}},\ \bibinfo {pages} {263} (\bibinfo {year} {1998})},\ \bibinfo {note}
  {hep-ph/9803315}\BibitemShut {NoStop}%
\bibitem [{\citenamefont {Antoniadis}\ \emph {et~al.}(1998)\citenamefont
  {Antoniadis}, \citenamefont {Arkani-Hamed}, \citenamefont {Dimopoulos},\ and\
  \citenamefont {Dvali}}]{Antoniadis:1998ig}%
  \BibitemOpen
  \bibfield  {author} {\bibinfo {author} {\bibfnamefont {I.}~\bibnamefont
  {Antoniadis}}, \bibinfo {author} {\bibfnamefont {N.}~\bibnamefont
  {Arkani-Hamed}}, \bibinfo {author} {\bibfnamefont {S.}~\bibnamefont
  {Dimopoulos}}, \ and\ \bibinfo {author} {\bibfnamefont {G.~R.}\ \bibnamefont
  {Dvali}},\ }\href {\doibase 10.1016/S0370-2693(98)00860-0} {\bibfield
  {journal} {\bibinfo  {journal} {Phys. Lett. B}\ }\textbf {\bibinfo {volume}
  {436}},\ \bibinfo {pages} {257} (\bibinfo {year} {1998})},\ \bibinfo {note}
  {hep-ph/9804398}\BibitemShut {NoStop}%
\bibitem [{\citenamefont {Randall}\ and\ \citenamefont
  {Sundrum}(1999{\natexlab{a}})}]{Randall:1999ee}%
  \BibitemOpen
  \bibfield  {author} {\bibinfo {author} {\bibfnamefont {L.}~\bibnamefont
  {Randall}}\ and\ \bibinfo {author} {\bibfnamefont {R.}~\bibnamefont
  {Sundrum}},\ }\href {\doibase 10.1103/PhysRevLett.83.3370} {\bibfield
  {journal} {\bibinfo  {journal} {Phys. Rev. Lett.}\ }\textbf {\bibinfo
  {volume} {83}},\ \bibinfo {pages} {3370} (\bibinfo {year}
  {1999}{\natexlab{a}})},\ \bibinfo {note} {hep-ph/9905221}\BibitemShut
  {NoStop}%
\bibitem [{\citenamefont {Randall}\ and\ \citenamefont
  {Sundrum}(1999{\natexlab{b}})}]{Randall:1999vf}%
  \BibitemOpen
  \bibfield  {author} {\bibinfo {author} {\bibfnamefont {L.}~\bibnamefont
  {Randall}}\ and\ \bibinfo {author} {\bibfnamefont {R.}~\bibnamefont
  {Sundrum}},\ }\href {\doibase 10.1103/PhysRevLett.83.4690} {\bibfield
  {journal} {\bibinfo  {journal} {Phys. Rev. Lett.}\ }\textbf {\bibinfo
  {volume} {83}},\ \bibinfo {pages} {4690} (\bibinfo {year}
  {1999}{\natexlab{b}})},\ \bibinfo {note} {hep-th/9906064}\BibitemShut
  {NoStop}%
\bibitem [{\citenamefont {Aaboud}\ \emph {et~al.}(2017)\citenamefont {Aaboud}
  \emph {et~al.}}]{Aaboud:2017yvp}%
  \BibitemOpen
  \bibfield  {author} {\bibinfo {author} {\bibfnamefont {M.}~\bibnamefont
  {Aaboud}} \emph {et~al.} (\bibinfo {collaboration} {ATLAS}),\ }\href@noop {}
  {\  (\bibinfo {year} {2017})},\ \Eprint {http://arxiv.org/abs/1703.09127}
  {arXiv:1703.09127 [hep-ex]} \BibitemShut {NoStop}%
%%CITATION = ARXIV:1703.09127;%%
\bibitem [{\citenamefont {Sirunyan}\ \emph {et~al.}(2017)\citenamefont
  {Sirunyan} \emph {et~al.}}]{Sirunyan:2017anm}%
  \BibitemOpen
  \bibfield  {author} {\bibinfo {author} {\bibfnamefont {A.~M.}\ \bibnamefont
  {Sirunyan}} \emph {et~al.} (\bibinfo {collaboration} {CMS}),\ }\href@noop {}
  {\  (\bibinfo {year} {2017})},\ \Eprint {http://arxiv.org/abs/1705.01403}
  {arXiv:1705.01403 [hep-ex]} \BibitemShut {NoStop}%
%%CITATION = ARXIV:1705.01403;%%
\bibitem [{\citenamefont {Cardoso}\ \emph {et~al.}(2015)\citenamefont
  {Cardoso}, \citenamefont {Gualtieri}, \citenamefont {Herdeiro},\ and\
  \citenamefont {Sperhake}}]{Cardoso:2014uka}%
  \BibitemOpen
  \bibfield  {author} {\bibinfo {author} {\bibfnamefont {V.}~\bibnamefont
  {Cardoso}}, \bibinfo {author} {\bibfnamefont {L.}~\bibnamefont {Gualtieri}},
  \bibinfo {author} {\bibfnamefont {C.}~\bibnamefont {Herdeiro}}, \ and\
  \bibinfo {author} {\bibfnamefont {U.}~\bibnamefont {Sperhake}},\ }\href
  {\doibase 10.1007/lrr-2015-1} {\bibfield  {journal} {\bibinfo  {journal}
  {Living Rev. Relativity}\ }\textbf {\bibinfo {volume} {18}},\ \bibinfo
  {pages} {1} (\bibinfo {year} {2015})},\ \bibinfo {note} {arXiv:1409.0014
  [gr-qc]}\BibitemShut {NoStop}%
\bibitem [{\citenamefont {Emparan}\ \emph {et~al.}(2013)\citenamefont
  {Emparan}, \citenamefont {Suzuki},\ and\ \citenamefont
  {Tanabe}}]{Emparan:2013moa}%
  \BibitemOpen
  \bibfield  {author} {\bibinfo {author} {\bibfnamefont {R.}~\bibnamefont
  {Emparan}}, \bibinfo {author} {\bibfnamefont {R.}~\bibnamefont {Suzuki}}, \
  and\ \bibinfo {author} {\bibfnamefont {K.}~\bibnamefont {Tanabe}},\ }\href
  {\doibase 10.1007/JHEP06(2013)009} {\bibfield  {journal} {\bibinfo  {journal}
  {JHEP}\ }\textbf {\bibinfo {volume} {06}},\ \bibinfo {pages} {009} (\bibinfo
  {year} {2013})},\ \bibinfo {note} {arXiv:1302.6382 [hep-th]}\BibitemShut
  {NoStop}%
\bibitem [{\citenamefont {Emparan}\ \emph
  {et~al.}(2015{\natexlab{a}})\citenamefont {Emparan}, \citenamefont
  {Shiromizu}, \citenamefont {Suzuki}, \citenamefont {Tanabe},\ and\
  \citenamefont {Tanaka}}]{Emparan:2015hwa}%
  \BibitemOpen
  \bibfield  {author} {\bibinfo {author} {\bibfnamefont {R.}~\bibnamefont
  {Emparan}}, \bibinfo {author} {\bibfnamefont {T.}~\bibnamefont {Shiromizu}},
  \bibinfo {author} {\bibfnamefont {R.}~\bibnamefont {Suzuki}}, \bibinfo
  {author} {\bibfnamefont {K.}~\bibnamefont {Tanabe}}, \ and\ \bibinfo {author}
  {\bibfnamefont {T.}~\bibnamefont {Tanaka}},\ }\href {\doibase
  10.1007/JHEP06(2015)159} {\bibfield  {journal} {\bibinfo  {journal} {JHEP}\
  }\textbf {\bibinfo {volume} {06}},\ \bibinfo {pages} {159} (\bibinfo {year}
  {2015}{\natexlab{a}})},\ \bibinfo {note} {arXiv:1504.06489
  [hep-th]}\BibitemShut {NoStop}%
\bibitem [{\citenamefont {Emparan}\ \emph
  {et~al.}(2015{\natexlab{b}})\citenamefont {Emparan}, \citenamefont {Suzuki},\
  and\ \citenamefont {Tanabe}}]{Emparan:2015gva}%
  \BibitemOpen
  \bibfield  {author} {\bibinfo {author} {\bibfnamefont {R.}~\bibnamefont
  {Emparan}}, \bibinfo {author} {\bibfnamefont {R.}~\bibnamefont {Suzuki}}, \
  and\ \bibinfo {author} {\bibfnamefont {K.}~\bibnamefont {Tanabe}},\ }\href
  {\doibase 10.1103/PhysRevLett.115.091102} {\bibfield  {journal} {\bibinfo
  {journal} {Phys. Rev. Lett.}\ }\textbf {\bibinfo {volume} {115}},\ \bibinfo
  {pages} {091102} (\bibinfo {year} {2015}{\natexlab{b}})},\ \Eprint
  {http://arxiv.org/abs/1506.06772} {arXiv:1506.06772 [hep-th]} \BibitemShut
  {NoStop}%
%%CITATION = ARXIV:1506.06772;%%
\bibitem [{\citenamefont {Emparan}\ and\ \citenamefont
  {Tanabe}(2014)}]{Emparan:2014cia}%
  \BibitemOpen
  \bibfield  {author} {\bibinfo {author} {\bibfnamefont {R.}~\bibnamefont
  {Emparan}}\ and\ \bibinfo {author} {\bibfnamefont {K.}~\bibnamefont
  {Tanabe}},\ }\href {\doibase 10.1103/PhysRevD.89.064028} {\bibfield
  {journal} {\bibinfo  {journal} {Phys. Rev.}\ }\textbf {\bibinfo {volume}
  {D89}},\ \bibinfo {pages} {064028} (\bibinfo {year} {2014})},\ \Eprint
  {http://arxiv.org/abs/1401.1957} {arXiv:1401.1957 [hep-th]} \BibitemShut
  {NoStop}%
%%CITATION = ARXIV:1401.1957;%%
\bibitem [{\citenamefont {Witek}\ \emph {et~al.}(2011)\citenamefont {Witek},
  \citenamefont {Cardoso}, \citenamefont {Gualtieri}, \citenamefont {Herdeiro},
  \citenamefont {Sperhake},\ and\ \citenamefont {Zilh{\~a}o}}]{Witek:2010az}%
  \BibitemOpen
  \bibfield  {author} {\bibinfo {author} {\bibfnamefont {H.}~\bibnamefont
  {Witek}}, \bibinfo {author} {\bibfnamefont {V.}~\bibnamefont {Cardoso}},
  \bibinfo {author} {\bibfnamefont {L.}~\bibnamefont {Gualtieri}}, \bibinfo
  {author} {\bibfnamefont {C.}~\bibnamefont {Herdeiro}}, \bibinfo {author}
  {\bibfnamefont {U.}~\bibnamefont {Sperhake}}, \ and\ \bibinfo {author}
  {\bibfnamefont {M.}~\bibnamefont {Zilh{\~a}o}},\ }\href {\doibase
  10.1103/PhysRevD.83.044017} {\bibfield  {journal} {\bibinfo  {journal} {Phys.
  Rev. D}\ }\textbf {\bibinfo {volume} {83}},\ \bibinfo {pages} {044017}
  (\bibinfo {year} {2011})},\ \bibinfo {note} {arXiv:1011.0742
  [gr-qc]}\BibitemShut {NoStop}%
\bibitem [{\citenamefont {Witek}\ \emph {et~al.}(2014)\citenamefont {Witek},
  \citenamefont {Okawa}, \citenamefont {Cardoso}, \citenamefont {Gualtieri},
  \citenamefont {Herdeiro}, \citenamefont {Shibata}, \citenamefont {Sperhake},\
  and\ \citenamefont {Zilh{\~a}o}}]{Witek:2014mha}%
  \BibitemOpen
  \bibfield  {author} {\bibinfo {author} {\bibfnamefont {H.}~\bibnamefont
  {Witek}}, \bibinfo {author} {\bibfnamefont {H.}~\bibnamefont {Okawa}},
  \bibinfo {author} {\bibfnamefont {V.}~\bibnamefont {Cardoso}}, \bibinfo
  {author} {\bibfnamefont {L.}~\bibnamefont {Gualtieri}}, \bibinfo {author}
  {\bibfnamefont {C.}~\bibnamefont {Herdeiro}}, \bibinfo {author}
  {\bibfnamefont {M.}~\bibnamefont {Shibata}}, \bibinfo {author} {\bibfnamefont
  {U.}~\bibnamefont {Sperhake}}, \ and\ \bibinfo {author} {\bibfnamefont
  {M.}~\bibnamefont {Zilh{\~a}o}},\ }\href {\doibase
  10.1103/PhysRevD.90.084014} {\bibfield  {journal} {\bibinfo  {journal} {Phys.
  Rev. D}\ }\textbf {\bibinfo {volume} {90}},\ \bibinfo {pages} {084014}
  (\bibinfo {year} {2014})},\ \bibinfo {note} {arXiv:1406.2703
  [gr-qc]}\BibitemShut {NoStop}%
\bibitem [{\citenamefont {Pretorius}(2005)}]{Pretorius:2004jg}%
  \BibitemOpen
  \bibfield  {author} {\bibinfo {author} {\bibfnamefont {F.}~\bibnamefont
  {Pretorius}},\ }\href {\doibase 10.1088/0264-9381/22/2/014} {\bibfield
  {journal} {\bibinfo  {journal} {Class. Quantum Grav.}\ }\textbf {\bibinfo
  {volume} {22}},\ \bibinfo {pages} {425} (\bibinfo {year} {2005})},\ \bibinfo
  {note} {gr-qc/0407110}\BibitemShut {NoStop}%
\bibitem [{\citenamefont {Yoshino}\ and\ \citenamefont
  {Shibata}(2009)}]{Yoshino:2009xp}%
  \BibitemOpen
  \bibfield  {author} {\bibinfo {author} {\bibfnamefont {H.}~\bibnamefont
  {Yoshino}}\ and\ \bibinfo {author} {\bibfnamefont {M.}~\bibnamefont
  {Shibata}},\ }\href {\doibase 10.1103/PhysRevD.80.084025} {\bibfield
  {journal} {\bibinfo  {journal} {Phys. Rev. D}\ }\textbf {\bibinfo {volume}
  {80}},\ \bibinfo {pages} {084025} (\bibinfo {year} {2009})},\ \bibinfo {note}
  {arXiv:0907.2760 [gr-qc]}\BibitemShut {NoStop}%
\bibitem [{\citenamefont {Cook}\ \emph {et~al.}(2016)\citenamefont {Cook},
  \citenamefont {Figueras}, \citenamefont {Kunesch}, \citenamefont {Sperhake},\
  and\ \citenamefont {Tunyasuvunakool}}]{Cook:2016soy}%
  \BibitemOpen
  \bibfield  {author} {\bibinfo {author} {\bibfnamefont {W.~G.}\ \bibnamefont
  {Cook}}, \bibinfo {author} {\bibfnamefont {P.}~\bibnamefont {Figueras}},
  \bibinfo {author} {\bibfnamefont {M.}~\bibnamefont {Kunesch}}, \bibinfo
  {author} {\bibfnamefont {U.}~\bibnamefont {Sperhake}}, \ and\ \bibinfo
  {author} {\bibfnamefont {S.}~\bibnamefont {Tunyasuvunakool}},\ }in\ \href
  {\doibase 10.1142/S0218271816410133} {\emph {\bibinfo {booktitle} {{3rd
  Amazonian Symposium on Physics and 5th NRHEP Network Meeting: Celebrating 100
  Years of General Relativity Belem, Brazil, September 28-October 2, 2015}}}},\
  Vol.~\bibinfo {volume} {25}\ (\bibinfo {year} {2016})\ p.\ \bibinfo {pages}
  {1641013},\ \bibinfo {note} {arXiv:1603.00362 [gr-qc]}\BibitemShut {NoStop}%
\bibitem [{\citenamefont {Cook}\ and\ \citenamefont
  {Sperhake}(2017)}]{Cook:2016qnt}%
  \BibitemOpen
  \bibfield  {author} {\bibinfo {author} {\bibfnamefont {W.~G.}\ \bibnamefont
  {Cook}}\ and\ \bibinfo {author} {\bibfnamefont {U.}~\bibnamefont
  {Sperhake}},\ }\href {\doibase 10.1088/1361-6382/aa5294} {\bibfield
  {journal} {\bibinfo  {journal} {Class. Quant. Grav.}\ }\textbf {\bibinfo
  {volume} {34}},\ \bibinfo {pages} {035010} (\bibinfo {year} {2017})},\
  \bibinfo {note} {arXiv:1609.01292 [gr-qc]}\BibitemShut {NoStop}%
\bibitem [{\citenamefont {Regge}\ and\ \citenamefont
  {Wheeler}(1957)}]{Regge:1957td}%
  \BibitemOpen
  \bibfield  {author} {\bibinfo {author} {\bibfnamefont {T.}~\bibnamefont
  {Regge}}\ and\ \bibinfo {author} {\bibfnamefont {J.~A.}\ \bibnamefont
  {Wheeler}},\ }\href {\doibase 10.1103/PhysRev.108.1063} {\bibfield  {journal}
  {\bibinfo  {journal} {Phys. Rev.}\ }\textbf {\bibinfo {volume} {108}},\
  \bibinfo {pages} {1063} (\bibinfo {year} {1957})}\BibitemShut {NoStop}%
\bibitem [{\citenamefont {Zerilli}(1970)}]{Zerilli:1971wd}%
  \BibitemOpen
  \bibfield  {author} {\bibinfo {author} {\bibfnamefont {F.~J.}\ \bibnamefont
  {Zerilli}},\ }\href {\doibase 10.1103/PhysRevD.2.2141} {\bibfield  {journal}
  {\bibinfo  {journal} {Phys. Rev. D}\ }\textbf {\bibinfo {volume} {2}},\
  \bibinfo {pages} {2141} (\bibinfo {year} {1970})}\BibitemShut {NoStop}%
\bibitem [{\citenamefont {Davis}\ \emph {et~al.}(1971)\citenamefont {Davis},
  \citenamefont {Ruffini}, \citenamefont {Press},\ and\ \citenamefont
  {Price}}]{Davis:1971gg}%
  \BibitemOpen
  \bibfield  {author} {\bibinfo {author} {\bibfnamefont {M.}~\bibnamefont
  {Davis}}, \bibinfo {author} {\bibfnamefont {R.}~\bibnamefont {Ruffini}},
  \bibinfo {author} {\bibfnamefont {W.~H.}\ \bibnamefont {Press}}, \ and\
  \bibinfo {author} {\bibfnamefont {R.~H.}\ \bibnamefont {Price}},\ }\href
  {\doibase 10.1103/PhysRevLett.27.1466} {\bibfield  {journal} {\bibinfo
  {journal} {Phys. Rev. Lett.}\ }\textbf {\bibinfo {volume} {27}},\ \bibinfo
  {pages} {1466} (\bibinfo {year} {1971})}\BibitemShut {NoStop}%
%%CITATION = PRLTA,27,1466;%%
\bibitem [{\citenamefont {Kodama}\ and\ \citenamefont
  {Ishibashi}(2003)}]{Kodama:2003jz}%
  \BibitemOpen
  \bibfield  {author} {\bibinfo {author} {\bibfnamefont {H.}~\bibnamefont
  {Kodama}}\ and\ \bibinfo {author} {\bibfnamefont {A.}~\bibnamefont
  {Ishibashi}},\ }\href {\doibase 10.1143/PTP.110.701} {\bibfield  {journal}
  {\bibinfo  {journal} {Prog. Theor. Phys.}\ }\textbf {\bibinfo {volume}
  {110}},\ \bibinfo {pages} {701} (\bibinfo {year} {2003})},\ \bibinfo {note}
  {hep-th/0305147}\BibitemShut {NoStop}%
\bibitem [{\citenamefont {Berti}\ \emph {et~al.}(2004)\citenamefont {Berti},
  \citenamefont {Cavaglia},\ and\ \citenamefont {Gualtieri}}]{Berti:2003si}%
  \BibitemOpen
  \bibfield  {author} {\bibinfo {author} {\bibfnamefont {E.}~\bibnamefont
  {Berti}}, \bibinfo {author} {\bibfnamefont {M.}~\bibnamefont {Cavaglia}}, \
  and\ \bibinfo {author} {\bibfnamefont {L.}~\bibnamefont {Gualtieri}},\ }\href
  {\doibase 10.1103/PhysRevD.69.124011} {\bibfield  {journal} {\bibinfo
  {journal} {Phys. Rev. D}\ }\textbf {\bibinfo {volume} {69}},\ \bibinfo
  {pages} {124011} (\bibinfo {year} {2004})},\ \bibinfo {note}
  {hep-th/0309203}\BibitemShut {NoStop}%
\bibitem [{\citenamefont {Berti}\ \emph {et~al.}(2011)\citenamefont {Berti},
  \citenamefont {Cardoso},\ and\ \citenamefont {Kipapa}}]{Berti:2010gx}%
  \BibitemOpen
  \bibfield  {author} {\bibinfo {author} {\bibfnamefont {E.}~\bibnamefont
  {Berti}}, \bibinfo {author} {\bibfnamefont {V.}~\bibnamefont {Cardoso}}, \
  and\ \bibinfo {author} {\bibfnamefont {B.}~\bibnamefont {Kipapa}},\ }\href
  {\doibase 10.1103/PhysRevD.83.084018} {\bibfield  {journal} {\bibinfo
  {journal} {Phys. Rev. D}\ }\textbf {\bibinfo {volume} {83}},\ \bibinfo
  {pages} {084018} (\bibinfo {year} {2011})}\BibitemShut {NoStop}%
\bibitem [{\citenamefont {Pretorius}(2009)}]{Pretorius:2007nq}%
  \BibitemOpen
  \bibfield  {author} {\bibinfo {author} {\bibfnamefont {F.}~\bibnamefont
  {Pretorius}},\ }in\ \href@noop {} {\emph {\bibinfo {booktitle} {{Physics of
  Relativistic Objects in Compact Binaries: From Birth to Coalescence}}}},\
  \bibinfo {editor} {edited by\ \bibinfo {editor} {\bibfnamefont
  {M.}~\bibnamefont {{Colpi {\em et al.}}}}}\ (\bibinfo  {publisher} {Springer,
  New York},\ \bibinfo {year} {2009})\ \bibinfo {note} {arXiv:0710.1338
  [gr-qc]}\BibitemShut {NoStop}%
\bibitem [{\citenamefont {Centrella}\ \emph {et~al.}(2010)\citenamefont
  {Centrella}, \citenamefont {Baker}, \citenamefont {Kelly},\ and\
  \citenamefont {van Meter}}]{Centrella:2010mx}%
  \BibitemOpen
  \bibfield  {author} {\bibinfo {author} {\bibfnamefont {J.~M.}\ \bibnamefont
  {Centrella}}, \bibinfo {author} {\bibfnamefont {J.~G.}\ \bibnamefont
  {Baker}}, \bibinfo {author} {\bibfnamefont {B.~J.}\ \bibnamefont {Kelly}}, \
  and\ \bibinfo {author} {\bibfnamefont {J.~R.}\ \bibnamefont {van Meter}},\
  }\href {\doibase 10.1103/RevModPhys.82.3069} {\bibfield  {journal} {\bibinfo
  {journal} {Rev. Mod. Phys.}\ }\textbf {\bibinfo {volume} {82}},\ \bibinfo
  {pages} {3069} (\bibinfo {year} {2010})},\ \bibinfo {note} {arXiv:1010.5260
  [gr-qc]}\BibitemShut {NoStop}%
\bibitem [{\citenamefont {Sperhake}(2015)}]{Sperhake:2014wpa}%
  \BibitemOpen
  \bibfield  {author} {\bibinfo {author} {\bibfnamefont {U.}~\bibnamefont
  {Sperhake}},\ }\href {\doibase 10.1088/0264-9381/32/12/124011} {\bibfield
  {journal} {\bibinfo  {journal} {Class. Quant. Grav.}\ }\textbf {\bibinfo
  {volume} {32}},\ \bibinfo {pages} {124011} (\bibinfo {year}
  {2015})}\BibitemShut {NoStop}%
\bibitem [{\citenamefont {Zilh{\~a}o}\ \emph {et~al.}(2010)\citenamefont
  {Zilh{\~a}o}, \citenamefont {Witek}, \citenamefont {Sperhake}, \citenamefont
  {Cardoso}, \citenamefont {Gualtieri}, \citenamefont {Herdeiro},\ and\
  \citenamefont {Nerozzi}}]{Zilhao:2010sr}%
  \BibitemOpen
  \bibfield  {author} {\bibinfo {author} {\bibfnamefont {M.}~\bibnamefont
  {Zilh{\~a}o}}, \bibinfo {author} {\bibfnamefont {H.}~\bibnamefont {Witek}},
  \bibinfo {author} {\bibfnamefont {U.}~\bibnamefont {Sperhake}}, \bibinfo
  {author} {\bibfnamefont {V.}~\bibnamefont {Cardoso}}, \bibinfo {author}
  {\bibfnamefont {L.}~\bibnamefont {Gualtieri}}, \bibinfo {author}
  {\bibfnamefont {C.}~\bibnamefont {Herdeiro}}, \ and\ \bibinfo {author}
  {\bibfnamefont {A.}~\bibnamefont {Nerozzi}},\ }\href {\doibase
  10.1103/PhysRevD.81.084052} {\bibfield  {journal} {\bibinfo  {journal} {Phys.
  Rev. D}\ }\textbf {\bibinfo {volume} {81}},\ \bibinfo {pages} {084052}
  (\bibinfo {year} {2010})},\ \bibinfo {note} {arXiv:1001.2302
  [gr-qc]}\BibitemShut {NoStop}%
\bibitem [{\citenamefont {Sorkin}(2010)}]{Sorkin:2009wh}%
  \BibitemOpen
  \bibfield  {author} {\bibinfo {author} {\bibfnamefont {E.}~\bibnamefont
  {Sorkin}},\ }\href {\doibase 10.1103/PhysRevD.81.084062} {\bibfield
  {journal} {\bibinfo  {journal} {Phys. Rev. D}\ }\textbf {\bibinfo {volume}
  {81}},\ \bibinfo {pages} {084062} (\bibinfo {year} {2010})},\ \bibinfo {note}
  {arXiv:0911.2011 [gr-qc]}\BibitemShut {NoStop}%
\bibitem [{\citenamefont {Yoshino}\ and\ \citenamefont
  {Shibata}(2011)}]{Yoshino:2011zz}%
  \BibitemOpen
  \bibfield  {author} {\bibinfo {author} {\bibfnamefont {H.}~\bibnamefont
  {Yoshino}}\ and\ \bibinfo {author} {\bibfnamefont {M.}~\bibnamefont
  {Shibata}},\ }\href {\doibase 10.1143/PTPS.189.269} {\bibfield  {journal}
  {\bibinfo  {journal} {Prog.Theor.Phys.Suppl.}\ }\textbf {\bibinfo {volume}
  {189}},\ \bibinfo {pages} {269} (\bibinfo {year} {2011})}\BibitemShut
  {NoStop}%
\bibitem [{\citenamefont {Sperhake}(2007)}]{Sperhake:2006cy}%
  \BibitemOpen
  \bibfield  {author} {\bibinfo {author} {\bibfnamefont {U.}~\bibnamefont
  {Sperhake}},\ }\href {\doibase 10.1103/PhysRevD.76.104015} {\bibfield
  {journal} {\bibinfo  {journal} {Phys. Rev. D}\ }\textbf {\bibinfo {volume}
  {76}},\ \bibinfo {pages} {104015} (\bibinfo {year} {2007})},\ \bibinfo {note}
  {gr-qc/0606079}\BibitemShut {NoStop}%
\bibitem [{\citenamefont {Sperhake}\ \emph {et~al.}(2008)\citenamefont
  {Sperhake}, \citenamefont {Berti}, \citenamefont {Cardoso}, \citenamefont
  {Gonz{\'a}lez}, \citenamefont {Br{\"u}gmann},\ and\ \citenamefont
  {Ansorg}}]{Sperhake:2007gu}%
  \BibitemOpen
  \bibfield  {author} {\bibinfo {author} {\bibfnamefont {U.}~\bibnamefont
  {Sperhake}}, \bibinfo {author} {\bibfnamefont {E.}~\bibnamefont {Berti}},
  \bibinfo {author} {\bibfnamefont {V.}~\bibnamefont {Cardoso}}, \bibinfo
  {author} {\bibfnamefont {J.~A.}\ \bibnamefont {Gonz{\'a}lez}}, \bibinfo
  {author} {\bibfnamefont {B.}~\bibnamefont {Br{\"u}gmann}}, \ and\ \bibinfo
  {author} {\bibfnamefont {M.}~\bibnamefont {Ansorg}},\ }\href {\doibase
  10.1103/PhysRevD.78.064069} {\bibfield  {journal} {\bibinfo  {journal} {Phys.
  Rev. D}\ }\textbf {\bibinfo {volume} {78}},\ \bibinfo {pages} {064069}
  (\bibinfo {year} {2008})},\ \bibinfo {note} {arXiv:0710.3823
  [gr-qc]}\BibitemShut {NoStop}%
\bibitem [{\citenamefont {Brill}\ and\ \citenamefont
  {Lindquist}(1963)}]{Brill:1963yv}%
  \BibitemOpen
  \bibfield  {author} {\bibinfo {author} {\bibfnamefont {D.~R.}\ \bibnamefont
  {Brill}}\ and\ \bibinfo {author} {\bibfnamefont {R.~W.}\ \bibnamefont
  {Lindquist}},\ }\href {\doibase 10.1103/PhysRev.131.471} {\bibfield
  {journal} {\bibinfo  {journal} {Phys. Rev.}\ }\textbf {\bibinfo {volume}
  {131}},\ \bibinfo {pages} {471} (\bibinfo {year} {1963})}\BibitemShut
  {NoStop}%
\bibitem [{\citenamefont {Arnowitt}\ \emph {et~al.}(1962)\citenamefont
  {Arnowitt}, \citenamefont {Deser},\ and\ \citenamefont
  {Misner}}]{Arnowitt:1962hi}%
  \BibitemOpen
  \bibfield  {author} {\bibinfo {author} {\bibfnamefont {R.}~\bibnamefont
  {Arnowitt}}, \bibinfo {author} {\bibfnamefont {S.}~\bibnamefont {Deser}}, \
  and\ \bibinfo {author} {\bibfnamefont {C.~W.}\ \bibnamefont {Misner}},\ }in\
  \href@noop {} {\emph {\bibinfo {booktitle} {{Gravitation an introduction to
  current research}}}},\ \bibinfo {editor} {edited by\ \bibinfo {editor}
  {\bibfnamefont {L.}~\bibnamefont {Witten}}}\ (\bibinfo  {publisher} {John
  Wiley, New York},\ \bibinfo {year} {1962})\ pp.\ \bibinfo {pages}
  {227--265},\ \bibinfo {note} {gr-qc/0405109}\BibitemShut {NoStop}%
\bibitem [{\citenamefont {Shibata}\ and\ \citenamefont
  {Nakamura}(1995)}]{Shibata:1995we}%
  \BibitemOpen
  \bibfield  {author} {\bibinfo {author} {\bibfnamefont {M.}~\bibnamefont
  {Shibata}}\ and\ \bibinfo {author} {\bibfnamefont {T.}~\bibnamefont
  {Nakamura}},\ }\href {\doibase 10.1103/PhysRevD.52.5428} {\bibfield
  {journal} {\bibinfo  {journal} {Phys. Rev. D}\ }\textbf {\bibinfo {volume}
  {52}},\ \bibinfo {pages} {5428} (\bibinfo {year} {1995})}\BibitemShut
  {NoStop}%
\bibitem [{\citenamefont {Baumgarte}\ and\ \citenamefont
  {Shapiro}(1998)}]{Baumgarte:1998te}%
  \BibitemOpen
  \bibfield  {author} {\bibinfo {author} {\bibfnamefont {T.~W.}\ \bibnamefont
  {Baumgarte}}\ and\ \bibinfo {author} {\bibfnamefont {S.~L.}\ \bibnamefont
  {Shapiro}},\ }\href {\doibase 10.1103/PhysRevD.59.024007} {\bibfield
  {journal} {\bibinfo  {journal} {Phys. Rev. D}\ }\textbf {\bibinfo {volume}
  {59}},\ \bibinfo {pages} {024007} (\bibinfo {year} {1998})},\ \bibinfo {note}
  {gr-qc/9810065}\BibitemShut {NoStop}%
\bibitem [{\citenamefont {Baker}\ \emph {et~al.}(2006)\citenamefont {Baker},
  \citenamefont {Centrella}, \citenamefont {Choi}, \citenamefont {Koppitz},\
  and\ \citenamefont {van Meter}}]{Baker:2005vv}%
  \BibitemOpen
  \bibfield  {author} {\bibinfo {author} {\bibfnamefont {J.~G.}\ \bibnamefont
  {Baker}}, \bibinfo {author} {\bibfnamefont {J.}~\bibnamefont {Centrella}},
  \bibinfo {author} {\bibfnamefont {D.-I.}\ \bibnamefont {Choi}}, \bibinfo
  {author} {\bibfnamefont {M.}~\bibnamefont {Koppitz}}, \ and\ \bibinfo
  {author} {\bibfnamefont {J.}~\bibnamefont {van Meter}},\ }\href {\doibase
  10.1103/PhysRevLett.96.111102} {\bibfield  {journal} {\bibinfo  {journal}
  {Phys. Rev. Lett.}\ }\textbf {\bibinfo {volume} {96}},\ \bibinfo {pages}
  {111102} (\bibinfo {year} {2006})},\ \bibinfo {note}
  {gr-qc/0511103}\BibitemShut {NoStop}%
\bibitem [{\citenamefont {Campanelli}\ \emph {et~al.}(2006)\citenamefont
  {Campanelli}, \citenamefont {Lousto}, \citenamefont {Marronetti},\ and\
  \citenamefont {Zlochower}}]{Campanelli:2005dd}%
  \BibitemOpen
  \bibfield  {author} {\bibinfo {author} {\bibfnamefont {M.}~\bibnamefont
  {Campanelli}}, \bibinfo {author} {\bibfnamefont {C.~O.}\ \bibnamefont
  {Lousto}}, \bibinfo {author} {\bibfnamefont {P.}~\bibnamefont {Marronetti}},
  \ and\ \bibinfo {author} {\bibfnamefont {Y.}~\bibnamefont {Zlochower}},\
  }\href {\doibase 10.1103/PhysRevLett.96.111101} {\bibfield  {journal}
  {\bibinfo  {journal} {Phys. Rev. Lett.}\ }\textbf {\bibinfo {volume} {96}},\
  \bibinfo {pages} {111101} (\bibinfo {year} {2006})},\ \bibinfo {note}
  {gr-qc/0511048}\BibitemShut {NoStop}%
\bibitem [{\citenamefont {Schnetter}\ \emph {et~al.}(2004)\citenamefont
  {Schnetter}, \citenamefont {Hawley},\ and\ \citenamefont
  {Hawke}}]{Schnetter:2003rb}%
  \BibitemOpen
  \bibfield  {author} {\bibinfo {author} {\bibfnamefont {E.}~\bibnamefont
  {Schnetter}}, \bibinfo {author} {\bibfnamefont {S.~H.}\ \bibnamefont
  {Hawley}}, \ and\ \bibinfo {author} {\bibfnamefont {I.}~\bibnamefont
  {Hawke}},\ }\href {\doibase 10.1088/0264-9381/21/6/014} {\bibfield  {journal}
  {\bibinfo  {journal} {Class. Quant. Grav.}\ }\textbf {\bibinfo {volume}
  {21}},\ \bibinfo {pages} {1465} (\bibinfo {year} {2004})},\ \bibinfo {note}
  {gr-qc/0310042}\BibitemShut {NoStop}%
\bibitem [{zzz002()}]{Carpetweb}%
  \BibitemOpen
  zzz002,\ \href@noop {} {}\bibinfo {note} {{Carpet Code homepage}: {\tt
  http://www.carpetcode.org/}}\BibitemShut {NoStop}%
\bibitem [{\citenamefont {{Allen, G. and Goodale, T. and Mass{\'o}, J. and
  Seidel, E.}}(1999)}]{Allen:1999}%
  \BibitemOpen
  \bibfield  {author} {\bibinfo {author} {\bibnamefont {{Allen, G. and Goodale,
  T. and Mass{\'o}, J. and Seidel, E.}}},\ }in\ \href@noop {} {\emph {\bibinfo
  {booktitle} {{Proceedings of Eighth IEEE International Symposium on High
  Performance Distributed Computing, HPDC-8, Redondo Beach, 1999}}}}\ (\bibinfo
   {publisher} {{IEEE Press}},\ \bibinfo {address} {{}},\ \bibinfo {year}
  {1999})\BibitemShut {NoStop}%
\bibitem [{zzz001()}]{Cactusweb}%
  \BibitemOpen
  zzz001,\ \href@noop {} {}\bibinfo {note} {{Cactus Computational Toolkit
  homepage:} {\tt http://www.cactuscode.org/}}\BibitemShut {NoStop}%
\bibitem [{\citenamefont {Godazgar}\ and\ \citenamefont
  {Reall}(2012)}]{Godazgar:2012zq}%
  \BibitemOpen
  \bibfield  {author} {\bibinfo {author} {\bibfnamefont {M.}~\bibnamefont
  {Godazgar}}\ and\ \bibinfo {author} {\bibfnamefont {H.~S.}\ \bibnamefont
  {Reall}},\ }\href {\doibase 10.1103/PhysRevD.85.084021} {\bibfield  {journal}
  {\bibinfo  {journal} {Phys. Rev. D}\ }\textbf {\bibinfo {volume} {85}},\
  \bibinfo {pages} {084021} (\bibinfo {year} {2012})},\ \bibinfo {note}
  {arXiv:1201.4373 [gr-qc]}\BibitemShut {NoStop}%
\bibitem [{\citenamefont {Cook}\ and\ \citenamefont
  {Sperhake}()}]{Cook:inprep}%
  \BibitemOpen
  \bibfield  {author} {\bibinfo {author} {\bibfnamefont {W.~G.}\ \bibnamefont
  {Cook}}\ and\ \bibinfo {author} {\bibfnamefont {U.}~\bibnamefont
  {Sperhake}},\ }\href@noop {} {\enquote {\bibinfo {title} {{In
  preparation}},}\ }\BibitemShut {NoStop}%
\bibitem [{\citenamefont {Ishibashi}\ and\ \citenamefont
  {Kodama}(2011)}]{Ishibashi:2011ws}%
  \BibitemOpen
  \bibfield  {author} {\bibinfo {author} {\bibfnamefont {A.}~\bibnamefont
  {Ishibashi}}\ and\ \bibinfo {author} {\bibfnamefont {H.}~\bibnamefont
  {Kodama}},\ }\href {\doibase 10.1143/PTPS.189.165} {\bibfield  {journal}
  {\bibinfo  {journal} {Prog. Theor. Phys. Suppl.}\ }\textbf {\bibinfo {volume}
  {189}},\ \bibinfo {pages} {165} (\bibinfo {year} {2011})},\ \bibinfo {note}
  {arXiv:1103.6148 [hep-th]}\BibitemShut {NoStop}%
\bibitem [{\citenamefont {Witek}\ \emph {et~al.}(2010)\citenamefont {Witek},
  \citenamefont {Zilh{\~a}o}, \citenamefont {Gualtieri}, \citenamefont
  {Cardoso}, \citenamefont {Herdeiro}, \citenamefont {Nerozzi},\ and\
  \citenamefont {Sperhake}}]{Witek:2010xi}%
  \BibitemOpen
  \bibfield  {author} {\bibinfo {author} {\bibfnamefont {H.}~\bibnamefont
  {Witek}}, \bibinfo {author} {\bibfnamefont {M.}~\bibnamefont {Zilh{\~a}o}},
  \bibinfo {author} {\bibfnamefont {L.}~\bibnamefont {Gualtieri}}, \bibinfo
  {author} {\bibfnamefont {V.}~\bibnamefont {Cardoso}}, \bibinfo {author}
  {\bibfnamefont {C.}~\bibnamefont {Herdeiro}}, \bibinfo {author}
  {\bibfnamefont {A.}~\bibnamefont {Nerozzi}}, \ and\ \bibinfo {author}
  {\bibfnamefont {U.}~\bibnamefont {Sperhake}},\ }\href {\doibase
  10.1103/PhysRevD.82.104014} {\bibfield  {journal} {\bibinfo  {journal} {Phys.
  Rev. D}\ }\textbf {\bibinfo {volume} {82}},\ \bibinfo {pages} {104014}
  (\bibinfo {year} {2010})},\ \bibinfo {note} {arXiv:1006.3081
  [gr-qc]}\BibitemShut {NoStop}%
\bibitem [{\citenamefont {Reisswig}\ \emph {et~al.}(2011)\citenamefont
  {Reisswig}, \citenamefont {Ott}, \citenamefont {Sperhake},\ and\
  \citenamefont {Schnetter}}]{Reisswig:2010cd}%
  \BibitemOpen
  \bibfield  {author} {\bibinfo {author} {\bibfnamefont {C.}~\bibnamefont
  {Reisswig}}, \bibinfo {author} {\bibfnamefont {C.~D.}\ \bibnamefont {Ott}},
  \bibinfo {author} {\bibfnamefont {U.}~\bibnamefont {Sperhake}}, \ and\
  \bibinfo {author} {\bibfnamefont {E.}~\bibnamefont {Schnetter}},\ }\href
  {\doibase 10.1103/PhysRevD.83.064008} {\bibfield  {journal} {\bibinfo
  {journal} {Phys. Rev. D}\ }\textbf {\bibinfo {volume} {83}},\ \bibinfo
  {pages} {064008} (\bibinfo {year} {2011})},\ \bibinfo {note} {arXiv:1012.0595
  [gr-qc]}\BibitemShut {NoStop}%
\bibitem [{\citenamefont {Emparan}\ \emph {et~al.}(2014)\citenamefont
  {Emparan}, \citenamefont {Suzuki},\ and\ \citenamefont
  {Tanabe}}]{Emparan:2014aba}%
  \BibitemOpen
  \bibfield  {author} {\bibinfo {author} {\bibfnamefont {R.}~\bibnamefont
  {Emparan}}, \bibinfo {author} {\bibfnamefont {R.}~\bibnamefont {Suzuki}}, \
  and\ \bibinfo {author} {\bibfnamefont {K.}~\bibnamefont {Tanabe}},\ }\href
  {\doibase 10.1007/JHEP07(2014)113} {\bibfield  {journal} {\bibinfo  {journal}
  {JHEP}\ }\textbf {\bibinfo {volume} {07}},\ \bibinfo {pages} {113} (\bibinfo
  {year} {2014})},\ \Eprint {http://arxiv.org/abs/1406.1258} {arXiv:1406.1258
  [hep-th]} \BibitemShut {NoStop}%
%%CITATION = ARXIV:1406.1258;%%
\bibitem [{\citenamefont {Emparan}\ \emph
  {et~al.}(2015{\natexlab{c}})\citenamefont {Emparan}, \citenamefont {Suzuki},\
  and\ \citenamefont {Tanabe}}]{Emparan:2015rva}%
  \BibitemOpen
  \bibfield  {author} {\bibinfo {author} {\bibfnamefont {R.}~\bibnamefont
  {Emparan}}, \bibinfo {author} {\bibfnamefont {R.}~\bibnamefont {Suzuki}}, \
  and\ \bibinfo {author} {\bibfnamefont {K.}~\bibnamefont {Tanabe}},\ }\href
  {\doibase 10.1007/JHEP04(2015)085} {\bibfield  {journal} {\bibinfo  {journal}
  {JHEP}\ }\textbf {\bibinfo {volume} {04}},\ \bibinfo {pages} {085} (\bibinfo
  {year} {2015}{\natexlab{c}})},\ \Eprint {http://arxiv.org/abs/1502.02820}
  {arXiv:1502.02820 [hep-th]} \BibitemShut {NoStop}%
%%CITATION = ARXIV:1502.02820;%%
\bibitem [{\citenamefont {Dias}\ \emph {et~al.}(2014)\citenamefont {Dias},
  \citenamefont {Hartnett},\ and\ \citenamefont {Santos}}]{Dias:2014eua}%
  \BibitemOpen
  \bibfield  {author} {\bibinfo {author} {\bibfnamefont {O.~J.~C.}\
  \bibnamefont {Dias}}, \bibinfo {author} {\bibfnamefont {G.~S.}\ \bibnamefont
  {Hartnett}}, \ and\ \bibinfo {author} {\bibfnamefont {J.~E.}\ \bibnamefont
  {Santos}},\ }\href {\doibase 10.1088/0264-9381/31/24/245011} {\bibfield
  {journal} {\bibinfo  {journal} {Class. Quant. Grav.}\ }\textbf {\bibinfo
  {volume} {31}},\ \bibinfo {pages} {245011} (\bibinfo {year} {2014})},\
  \bibinfo {note} {arXiv:1402.7047 [hep-th]}\BibitemShut {NoStop}%
\bibitem [{\citenamefont {Cardoso}\ \emph {et~al.}(2003)\citenamefont
  {Cardoso}, \citenamefont {Dias},\ and\ \citenamefont
  {Lemos}}]{Cardoso:2002pa}%
  \BibitemOpen
  \bibfield  {author} {\bibinfo {author} {\bibfnamefont {V.}~\bibnamefont
  {Cardoso}}, \bibinfo {author} {\bibfnamefont {O.~J.~C.}\ \bibnamefont
  {Dias}}, \ and\ \bibinfo {author} {\bibfnamefont {J.~P.~S.}\ \bibnamefont
  {Lemos}},\ }\href {\doibase 10.1103/PhysRevD.67.064026} {\bibfield  {journal}
  {\bibinfo  {journal} {Phys. Rev. D}\ }\textbf {\bibinfo {volume} {67}},\
  \bibinfo {pages} {064026} (\bibinfo {year} {2003})},\ \bibinfo {note}
  {hep-th/0212168}\BibitemShut {NoStop}%
\bibitem [{\citenamefont {Berti}\ \emph {et~al.}(2007)\citenamefont {Berti},
  \citenamefont {Cardoso}, \citenamefont {Gonz{\'a}lez}, \citenamefont
  {Sperhake}, \citenamefont {Hannam}, \citenamefont {Husa},\ and\ \citenamefont
  {Br{\"u}gmann}}]{Berti:2007fi}%
  \BibitemOpen
  \bibfield  {author} {\bibinfo {author} {\bibfnamefont {E.}~\bibnamefont
  {Berti}}, \bibinfo {author} {\bibfnamefont {V.}~\bibnamefont {Cardoso}},
  \bibinfo {author} {\bibfnamefont {J.~A.}\ \bibnamefont {Gonz{\'a}lez}},
  \bibinfo {author} {\bibfnamefont {U.}~\bibnamefont {Sperhake}}, \bibinfo
  {author} {\bibfnamefont {M.~D.}\ \bibnamefont {Hannam}}, \bibinfo {author}
  {\bibfnamefont {S.}~\bibnamefont {Husa}}, \ and\ \bibinfo {author}
  {\bibfnamefont {B.}~\bibnamefont {Br{\"u}gmann}},\ }\href {\doibase
  10.1103/PhysRevD.76.064034} {\bibfield  {journal} {\bibinfo  {journal} {Phys.
  Rev. D}\ }\textbf {\bibinfo {volume} {76}},\ \bibinfo {pages} {064034}
  (\bibinfo {year} {2007})},\ \bibinfo {note} {gr-qc/0703053}\BibitemShut
  {NoStop}%
\bibitem [{\citenamefont {Le~Tiec}\ \emph {et~al.}(2011)\citenamefont
  {Le~Tiec}, \citenamefont {Mroue}, \citenamefont {Barack}, \citenamefont
  {Buonanno}, \citenamefont {Pfeiffer}, \citenamefont {Sago},\ and\
  \citenamefont {Taracchini}}]{LeTiec:2011bk}%
  \BibitemOpen
  \bibfield  {author} {\bibinfo {author} {\bibfnamefont {A.}~\bibnamefont
  {Le~Tiec}}, \bibinfo {author} {\bibfnamefont {A.~H.}\ \bibnamefont {Mroue}},
  \bibinfo {author} {\bibfnamefont {L.}~\bibnamefont {Barack}}, \bibinfo
  {author} {\bibfnamefont {A.}~\bibnamefont {Buonanno}}, \bibinfo {author}
  {\bibfnamefont {H.~P.}\ \bibnamefont {Pfeiffer}}, \bibinfo {author}
  {\bibfnamefont {N.}~\bibnamefont {Sago}}, \ and\ \bibinfo {author}
  {\bibfnamefont {A.}~\bibnamefont {Taracchini}},\ }\href {\doibase
  10.1103/PhysRevLett.107.141101} {\bibfield  {journal} {\bibinfo  {journal}
  {Phys. Rev. Lett.}\ }\textbf {\bibinfo {volume} {107}},\ \bibinfo {pages}
  {141101} (\bibinfo {year} {2011})},\ \Eprint {http://arxiv.org/abs/1106.3278}
  {arXiv:1106.3278 [gr-qc]} \BibitemShut {NoStop}%
%%CITATION = ARXIV:1106.3278;%%
\bibitem [{\citenamefont {{Le Tiec}}(2014)}]{Tiec:2014lba}%
  \BibitemOpen
  \bibfield  {author} {\bibinfo {author} {\bibfnamefont {A.}~\bibnamefont {{Le
  Tiec}}},\ }\href {\doibase 10.1142/S0218271814300225} {\bibfield  {journal}
  {\bibinfo  {journal} {Int. J. Mod. Phys. D}\ }\textbf {\bibinfo {volume}
  {23}},\ \bibinfo {pages} {1430022} (\bibinfo {year} {2014})},\ \bibinfo
  {note} {arXiv:1408.5505 [gr-qc]}\BibitemShut {NoStop}%
\bibitem [{\citenamefont {Nakamura}\ and\ \citenamefont
  {Haugan}(1983)}]{Nakamura:1983hk}%
  \BibitemOpen
  \bibfield  {author} {\bibinfo {author} {\bibfnamefont {T.}~\bibnamefont
  {Nakamura}}\ and\ \bibinfo {author} {\bibfnamefont {M.~P.}\ \bibnamefont
  {Haugan}},\ }\href {\doibase 10.1086/161041} {\bibfield  {journal} {\bibinfo
  {journal} {Astrophys. J.}\ }\textbf {\bibinfo {volume} {269}},\ \bibinfo
  {pages} {292} (\bibinfo {year} {1983})}\BibitemShut {NoStop}%
\bibitem [{\citenamefont {Berti}\ \emph {et~al.}(2010)\citenamefont {Berti},
  \citenamefont {Cardoso}, \citenamefont {Hinderer}, \citenamefont {Lemos},
  \citenamefont {Pretorius}, \citenamefont {Sperhake},\ and\ \citenamefont
  {Yunes}}]{Berti:2010ce}%
  \BibitemOpen
  \bibfield  {author} {\bibinfo {author} {\bibfnamefont {E.}~\bibnamefont
  {Berti}}, \bibinfo {author} {\bibfnamefont {V.}~\bibnamefont {Cardoso}},
  \bibinfo {author} {\bibfnamefont {T.}~\bibnamefont {Hinderer}}, \bibinfo
  {author} {\bibfnamefont {M.}~\bibnamefont {Lemos}}, \bibinfo {author}
  {\bibfnamefont {F.}~\bibnamefont {Pretorius}}, \bibinfo {author}
  {\bibfnamefont {U.}~\bibnamefont {Sperhake}}, \ and\ \bibinfo {author}
  {\bibfnamefont {N.}~\bibnamefont {Yunes}},\ }\href {\doibase
  10.1103/PhysRevD.81.104048} {\bibfield  {journal} {\bibinfo  {journal} {Phys.
  Rev. D}\ }\textbf {\bibinfo {volume} {81}},\ \bibinfo {pages} {104048}
  (\bibinfo {year} {2010})},\ \bibinfo {note} {arXiv:1003.0812
  [gr-qc]}\BibitemShut {NoStop}%
\bibitem [{\citenamefont {Sperhake}\ \emph {et~al.}(2011)\citenamefont
  {Sperhake}, \citenamefont {Cardoso}, \citenamefont {Ott}, \citenamefont
  {Schnetter},\ and\ \citenamefont {Witek}}]{Sperhake:2011ik}%
  \BibitemOpen
  \bibfield  {author} {\bibinfo {author} {\bibfnamefont {U.}~\bibnamefont
  {Sperhake}}, \bibinfo {author} {\bibfnamefont {V.}~\bibnamefont {Cardoso}},
  \bibinfo {author} {\bibfnamefont {C.~D.}\ \bibnamefont {Ott}}, \bibinfo
  {author} {\bibfnamefont {E.}~\bibnamefont {Schnetter}}, \ and\ \bibinfo
  {author} {\bibfnamefont {H.}~\bibnamefont {Witek}},\ }\href {\doibase
  10.1103/PhysRevD.84.084038} {\bibfield  {journal} {\bibinfo  {journal} {Phys.
  Rev. D}\ }\textbf {\bibinfo {volume} {84}},\ \bibinfo {pages} {084038}
  (\bibinfo {year} {2011})},\ \bibinfo {note} {arXiv:1105.5391
  [gr-qc]}\BibitemShut {NoStop}%
\bibitem [{\citenamefont {Fitchett}\ and\ \citenamefont
  {Detweiler}(1984)}]{Fitchett:1984qn}%
  \BibitemOpen
  \bibfield  {author} {\bibinfo {author} {\bibfnamefont {M.~J.}\ \bibnamefont
  {Fitchett}}\ and\ \bibinfo {author} {\bibfnamefont {S.}~\bibnamefont
  {Detweiler}},\ }\href@noop {} {\bibfield  {journal} {\bibinfo  {journal}
  {MNRAS}\ }\textbf {\bibinfo {volume} {211}},\ \bibinfo {pages} {933}
  (\bibinfo {year} {1984})}\BibitemShut {NoStop}%
\bibitem [{\citenamefont {Sperhake}\ \emph {et~al.}(2013)\citenamefont
  {Sperhake}, \citenamefont {Berti}, \citenamefont {Cardoso},\ and\
  \citenamefont {Pretorius}}]{Sperhake:2012me}%
  \BibitemOpen
  \bibfield  {author} {\bibinfo {author} {\bibfnamefont {U.}~\bibnamefont
  {Sperhake}}, \bibinfo {author} {\bibfnamefont {E.}~\bibnamefont {Berti}},
  \bibinfo {author} {\bibfnamefont {V.}~\bibnamefont {Cardoso}}, \ and\
  \bibinfo {author} {\bibfnamefont {F.}~\bibnamefont {Pretorius}},\ }\href
  {\doibase 10.1103/PhysRevLett.111.041101} {\bibfield  {journal} {\bibinfo
  {journal} {Phys. Rev. Lett.}\ }\textbf {\bibinfo {volume} {111}},\ \bibinfo
  {pages} {041101} (\bibinfo {year} {2013})},\ \bibinfo {note} {arXiv:1211.6114
  [gr-qc]}\BibitemShut {NoStop}%
\bibitem [{\citenamefont {Sperhake}\ \emph {et~al.}(2016)\citenamefont
  {Sperhake}, \citenamefont {Berti}, \citenamefont {Cardoso},\ and\
  \citenamefont {Pretorius}}]{Sperhake:2015siy}%
  \BibitemOpen
  \bibfield  {author} {\bibinfo {author} {\bibfnamefont {U.}~\bibnamefont
  {Sperhake}}, \bibinfo {author} {\bibfnamefont {E.}~\bibnamefont {Berti}},
  \bibinfo {author} {\bibfnamefont {V.}~\bibnamefont {Cardoso}}, \ and\
  \bibinfo {author} {\bibfnamefont {F.}~\bibnamefont {Pretorius}},\ }\href
  {\doibase 10.1103/PhysRevD.93.044012} {\bibfield  {journal} {\bibinfo
  {journal} {Phys. Rev. D}\ }\textbf {\bibinfo {volume} {93}},\ \bibinfo
  {pages} {044012} (\bibinfo {year} {2016})},\ \bibinfo {note}
  {arXiv:1511.08209 [gr-qc]}\BibitemShut {NoStop}%
\end{thebibliography}
%%\bibliographystyle{unsrt}


\end{document}